\documentclass[fleqn,usenatbib,useAMS]{mnras}

\usepackage{mathptmx}

\usepackage[T1]{fontenc}

\usepackage{graphicx}	
\usepackage{amsmath}	
\usepackage{amssymb}	
\usepackage{bm}

\DeclareMathAlphabet\mathbfcal{OMS}{cmsy}{b}{n}

\newcommand{\rxte}{\textit{RXTE}}

\newcommand{\fouru}{4U~1636-53}
\newcommand{\foru}{4U~1820-30}

\title[Disc Warps Driven by X-ray Bursts]{Radiation Driven Warping of
  Accretion Discs Due to X-ray Bursts}

\author[D. R. Ballantyne]{
D. R. Ballantyne\thanks{E-mail: david.ballantyne@physics.gatech.edu}
\\
Center for Relativistic Astrophysics, School of Physics, Georgia
  Institute of Technology, 837 State Street, Atlanta, GA 30332-0430, USA\\
}

\date{Accepted XXX. Received YYY; in original form ZZZ}

\pubyear{2022}

\begin{document}
\label{firstpage}
\pagerange{\pageref{firstpage}--\pageref{lastpage}}
\maketitle

\begin{abstract}
The outpouring of radiation during an X-ray burst can affect
the properties of accretion discs around neutron stars: the corona can
cool and collapse, the inner regions can be bled away due to enhanced
accretion, and the additional heating will lead to changes in the
disc height. In this paper, we investigate whether radiation
from bursts can cause the disc to distort through a warping
instability. Working in the limit of isotropic viscosity and linear
growth, we find that bursts are more likely to drive disc warps when they
have larger luminosities and longer durations. Therefore, warps will
be most probable during intermediate duration bursts (IMDBs) and
superbursts with evidence for photospheric radius expansion. Further, the
development of warps depends on the disc viscosity with larger
values of $\alpha$ increasing the likelihood of warp growth. We perform time-dependent
evolution calculations of the development of warps during Type I
bursts and IMDBs. Depending on the initial warp prior to the burst, we
find the burst produces warps at $r \la 50$~$r_g$ that rapidly grow and decay on
second-long timescales, or ones that grow more slowly and cover a
large fraction of the disc. The pulsations of warp at small radii appear to have the
properties needed to explain the achromatic fluctuations that have been
observed during the tails of some IMDBs. The large scale, slowly
growing warps could account for the large reflection strengths and
absorbing column densities inferred late in the \foru\ and
\fouru\ superbursts. 

\end{abstract}

\begin{keywords}
accretion, accretion discs -- hydrodynamics -- stars: neutron -- X-rays: binaries --
X-rays: bursts
\end{keywords}



\section{Introduction}
\label{sect:intro}
X-ray bursts are produced by unstable nuclear burning of gas accreted
onto the surfaces of weakly magnetic neutron stars \citep[e.g.,][]{lvt93,gk21}. The heat
produced by the nuclear reactions is thermalized and radiated as
X-rays by the stellar atmosphere \citep[e.g.,][]{spw11,spw12}. Depending on the nature of the
burning layer and the properties of the heat transfer through the
atmosphere, the timescale of the burst may be $\sim 10$~s
(known as a Type I burst), $\sim 100$~s (an intermediate duration
burst; IMDB), or, on rare occasions, $\sim 1000$~s (a
superburst) \citep[e.g.,][]{sb03}. Many thousands of bursts have been detected from over 100
galactic neutron stars (NSs), with the vast majority being the shorter Type
I bursts \citep{gall08,minbar20}.

The lightcurve of most bursts follows the rapid rise and slower decay
seen from explosive transients, although bursts may also exhibit features
such as plateaus or multiple peaks in their lightcurves \citep[e.g.,][]{igb11,barriere15,bult19,jai19,guver21,pike21,guver22}.
The peak luminosity often approaches, or even exceeds, the
Eddington luminosity of the neutron star ($\approx 2.5\times
10^{38} (M/M_{\odot})$~erg~s$^{-1}$, where $M$ is the mass of the
NS). Therefore, bursts are a powerful source of X-ray photons that are
explosively released into the stellar environment. Indeed,
observational evidence indicates that bursts can couple to the
surrounding accretion flow, significantly impacting both the disc and
corona \citep[e.g.,][]{be05,degenaar18}. For example, analysis of X-ray reflection features from
the disc during two long superbursts showed evidence for structural
changes in the disc on the timescale of the burst \citep{bs04,keek14b}. The
`persistent' accretion-powered X-ray emission is frequently observed
to be enhanced during bursts \citep[e.g.][]{worpel13,worpel15,guver22,zhao22}, which may result from an increase in
the local accretion rate due to Poynting-Robertson drag on the inner
disc from burst photons \citep{walker89,walker92}. In addition, hard X-ray deficits
observed during bursts would naturally occur from cooling of the
corona due to the burst of soft photons emanating from the NS
\citep{mc03,ji14,sanchez20,speicher20,chen22}. Recent numerical investigations of the interaction of bursts
with accretion flows confirm that X-ray bursts will significantly
affect the disc structure and environment in ways consistent with the
observational signatures \citep{fbmw18,fbb19}. These results suggest that X-ray bursts are
an important probe of accretion disc physics around NSs.

While perhaps the most dramatic examples of burst-disc interactions
occurs near the peak luminosity (e.g., the draining of the inner disc
due to Poynting-Robertson drag; \citealt{fbb19}), there may be long-lived, or
slowly growing, effects that only become significant during the fading
tails of bursts. For example, rapid fluctuations in the light-curve
have been seen during the decay of a small number of luminous bursts,
particularly IMDBs \citep[e.g.,][]{zand05,igb11,degenaar13,barriere15}. These fluctuations in brightness go above
and below the regular decay of the burst and have a typical timescale
of $\sim 1$~s. After a minute or two these fluctuations vanish as
quickly as they appeared. Another
example is found in the reflection analysis of the \fouru\ and
\foru\ superbursts. In both cases, time-resolved modeling of the
\rxte\ spectra found larger reflection fractions and obscuring column
densities during the tail of the burst than at the peak of the
burst \citep{bs04,keek14b}. Taken together, these results suggest that X-ray bursts,
especially the longer IMDBs and superbursts, may cause structural
changes to the disc that persist long after the peak luminosity.

Accretion discs illuminated by a strong central radiation source are
subject to a warping instability \citep{mbp96,pringle96,mbn98,od01}. Depending on the nature of
the illumination and viscous properties of the disc, the non-linear growth of the
warp can lead to strongly distorted discs that may precess on long
time-scales \citep{pringle97,wp99}. This effect has been frequently invoked to explain
long duration periodicities observed in some X-ray binaries
\citep[e.g.,][]{sood07}, and, as a result, previous work on radiation driven warping of discs focused on
the effects from steady, accretion-powered illumination \citep[e.g.,][]{od01}.
However, it is interesting to consider whether a transient, but
powerful, release of central radiation may also produce a temporary
disc warp. If an unstable disc warp can be driven by an
X-ray burst, then this may be a possible explanation for some of the
interesting behavior seen in the tails of long duration bursts
described above, and could comprise another probe of accretion physics
using X-ray bursts.

This paper investigates two questions related to the warping of
accretion discs by X-ray bursts. In Section~\ref{sect:criteria} we
evaluate the conditions necessary for a burst of a given luminosity
and duration to warp the accretion disc, and estimate the radius of
instability. In Section~\ref{sect:evolution} we solve the evolution
equation for a warped accretion disc to demonstrate the growth of the warping instability during bursts of
different durations. A discussion of the results is presented in
Sect.~\ref{sect:discuss} while Sect.~\ref{sect:concl} contains our
conclusions.

\section{Criteria for the Disc Warping Instability During an X-ray Burst}
\label{sect:criteria}

\subsection{The Critical Radius}
\label{sub:rcrit}
\citet{pringle97} compared the timescale for an accretion disc to warp due to
illumination by luminosity $L_{\ast}$ to the one for a disc to locally
flatten itself, and found that the warping instability occurs for disc radii $\geq
r_{\mathrm{crit}}$ where
\begin{equation}
  \label{eq:rcrit1}
  r_{\mathrm{crit}}=\left ({6 \pi c^2 \over L_{\ast}} \eta {\dot{M}
    \over 3 \pi} \gamma_{\mathrm{crit}}^{-1} \right )^2.
\end{equation}
Here, $r_{\mathrm{crit}}=R/r_g$ is the radius of the disc in units of gravitational radii
($r_g=GM/c^2$, where $M_{\ast}$ is the mass of the NS), $\dot{M}$ is the
accretion rate onto the NS, $\gamma_{\mathrm{crit}}$ is a numerical
factor determined by \citet{pringle97} to be $\approx 0.32$, and
$\eta=\nu_2/\nu_1$ (the ratio of the vertical and azimuthal shear
viscosities). This equation is valid for radii far from the inner
boundary, so that $\nu_1 \Sigma = \dot{M}/3\pi$ where $\Sigma$ is the
surface density of the disc.

In the case of an X-ray burst $L_{\ast} \approx L_b$, the burst
luminosity, as we are most interested in the case where $L_b \gg L_{\mathrm{acc}}$,
the overall accretion luminosity of the system. However, it is
convenient to scale $L_b$ by $L_{\mathrm{acc}}$ in order to connect to the accretion
rate of the disc. Thus,
\begin{equation}
  \label{eq:rcrit2}
  r_{\mathrm{crit}}=\left ({6 \pi c^2 \over L_{\mathrm{acc}} (L_b/L_{\mathrm{acc}})} \eta {\dot{M} \over 3 \pi} \gamma_{\mathrm{crit}}^{-1} \right )^2,
\end{equation}
but, since
\begin{equation}
  \label{eq:lumin}
 L_{\mathrm{acc}}={G M_{\ast} \dot{M} \over R_{\ast}},
\end{equation}
where $R_{\ast}$ is the radius of the NS, we
can eliminate $\dot{M}$ and $L_{\mathrm{acc}}$ from Eq.~\ref{eq:rcrit2} leaving
\begin{equation}
  \label{eq:rcrit3}
  r_{\mathrm{crit}}=\left ({6 \pi c^2 \over (L_b/L_{\mathrm{acc}})} \eta {R_{\ast}
    \over 3 \pi GM_{\ast}} \gamma_{\mathrm{crit}}^{-1} \right )^2,
\end{equation}
Finally, $M_{\ast}$ and $R_{\ast}$ can be replaced by the
gravitational redshift factor at the surface of a NS \citep{lvt93},
\begin{equation}
  \label{eq:zfactor}
  1+z_{\ast} = \left (1 - {2GM_{\ast} \over R_{\ast}c^2} \right
  )^{-1/2},
  \end{equation}
yielding our final expression for the critical radius for the warping
instability due to an X-ray burst,
\begin{equation}
\label{eq:rcrit}
r_{\mathrm{crit}} = \left ( \frac{4\eta} {\gamma_{\mathrm{crit}} \left ( 1-(1+z_{\ast})^{-2} \right ) (L_b/L_{\mathrm{acc}})} \right )^2.
\end{equation}
Thus, given a prescription for $\eta$, the values of
$r_{\mathrm{crit}}$ can be mapped out in the $(z_{\ast}, L_b/L_{\mathrm{acc}})$
plane.

\subsection{The Minimum Burst Timescale}
\label{sub:bursttime}
The critical radius criterion discussed above originates from comparing
the timescale of the growth of a warp, $t_{\mathrm{warp}}$, to the
timescale for the disc to flatten itself locally \citep{pringle97}. However, for an
X-ray burst to drive a warp, the warping timescale must be faster than
the burst duration, $t_{\mathrm{burst}}$. That is, in addition to
determining the radii at which the disc becomes unstable, we must
require $t_{\mathrm{warp}} < t_{\mathrm{burst}}$ at those radii.

Following \citet{pringle96,pringle97} this inequality can be written as
\begin{equation}
  \label{eq:tburst1}
  t_{\mathrm{burst}} > {12 \pi \Sigma R^3 \Omega c \over L_{\mathrm{acc}}(L_b/L_{\mathrm{acc}})},
\end{equation}
where $\Omega$ is the angular velocity of the disc. For a thin,
Keplerian disc far from the inner edge, we find
\begin{equation}
  \label{eq:tburst2}
  t_{\mathrm{burst}} > {12 \pi \over \nu_1}{\dot{M} \over 3\pi}{R^3
    c_s c \over HL_{\mathrm{acc}}(L_b/L_{\mathrm{acc}})},
\end{equation}
where $c_s$ is the sound speed and $H$ is the disc scale
height. Assuming $\nu_1 = \alpha c_s H$ \citep{ss73} and using
Eq.~\ref{eq:lumin} to eliminate $\dot{M}$ and $L_{\mathrm{acc}}$ yields
\begin{equation}
  \label{eq:tburst3}
  t_{\mathrm{burst}} > {4 \over \alpha}{R_{\ast} \over GM_{\ast}}{R^3
    c \over (L_b/L_{\mathrm{acc}})H^2}.
\end{equation}
The scale height of a gas-pressure dominated accretion disc with $e^-$
scattering opacity far from the inner boundary can be written as \citep[e.g.,][]{ss73}
\begin{equation}
  \label{eq:H1}
  H=(2.2\times 10^3 \mathrm{cm}) \alpha^{-1/10} \left ( {M_{\ast}
    \over M_{\odot}} \right )^{9/10} \left ( {\dot{M} \over
    \dot{M}_{\mathrm{Edd}}} \right )^{1/5} r^{21/20},
\end{equation}
where $\dot{M}_{\mathrm{Edd}}$ is the Eddington accretion rate and we
have assumed that the radiative efficiency of the disc is $0.1$\footnote{This
expression for $H$ is likely an underestimate during an X-ray burst,
as the disc will inflate due to X-ray heating \citep{fbb19}.}. As the dependencies on
$\alpha$ and $\dot{M}$ are relatively weak, the scale height can be
approximated as
\begin{equation}
  \label{eq:H2}
  H \approx (2.2\times 10^3 \mathrm{cm}) \left ( {M_{\ast}
    \over M_{\odot}} \right ) r.
\end{equation}
Substituting this expression into Eq.~\ref{eq:tburst3}, while also
making use of Eq.~\ref{eq:zfactor} to eliminate the stellar radius,
gives
\begin{equation}
  \label{eq:tburst}
  t_{\mathrm{burst}} \ga (0.17~\mathrm{s}) \left ( {
    (M_{\ast}/M_{\odot}) r \over \alpha (1-(1+z_{\ast})^{-2}) (L_b/L_{\mathrm{acc}})}
  \right ).
\end{equation}
Finally, evaluating the above expression at $r_{\mathrm{crit}}$ using
Eq.~\ref{eq:rcrit} gives the minimum burst duration needed to drive a
disc warp at $r_{\mathrm{crit}}$:
\begin{equation}
  \label{eq:tburstatrcrit}
  t_{\mathrm{burst}} \ga (2.8~\mathrm{s}) \left ( {
    (M_{\ast}/M_{\odot}) \eta^2 \over \alpha \gamma_{\mathrm{crit}}^2
    (1-(1+z_{\ast})^{-2})^3 (L_b/L_{\mathrm{acc}})^3} \right ).
\end{equation}

\subsection{Results}
\label{sub:res1}
We can now use Eqs.~\ref{eq:rcrit} and~\ref{eq:tburstatrcrit} to
estimate the burst luminosities and durations needed to drive a warp
at a particular radius on an accretion disc. However, before doing so,
the ratio of the vertical and azimuthal shear viscosities, $\eta$,
must first be defined.

Following the standard \citet{ss73} assumption, the vertical shear
viscosity is $\nu_2 = \alpha_2 c_s H$, and, as mentioned in
Sect.~\ref{sub:bursttime}, $\nu_1 = \alpha c_s H$, where both
$\alpha_2$ and $\alpha$ are treated as constants. For isotropic
viscosity and warps growing in the linear regime, $\alpha_2 = 1/2\alpha$ \citep{pp83,nixon15}, and therefore
\begin{equation}
  \label{eq:eta}
  \eta \approx {1 \over 2\alpha^2}.
\end{equation}
This isotropic approximation for $\eta$ is used for the remainder of
the paper\footnote{See \citet{ogilvie99} for a discussion
on the treatment of non-linear effects on the viscosity and the
implications for warped discs.}. Since the timescale for the vertical
viscosity to smooth out a disc warp is proportional to $\nu_2^{-1}$
\citep{pringle97}, a larger $\alpha$ will increase this timescale and make it
easier for radiation to drive a warp in an accretion disc.

\begin{figure*}
  \includegraphics[width=0.95\textwidth]{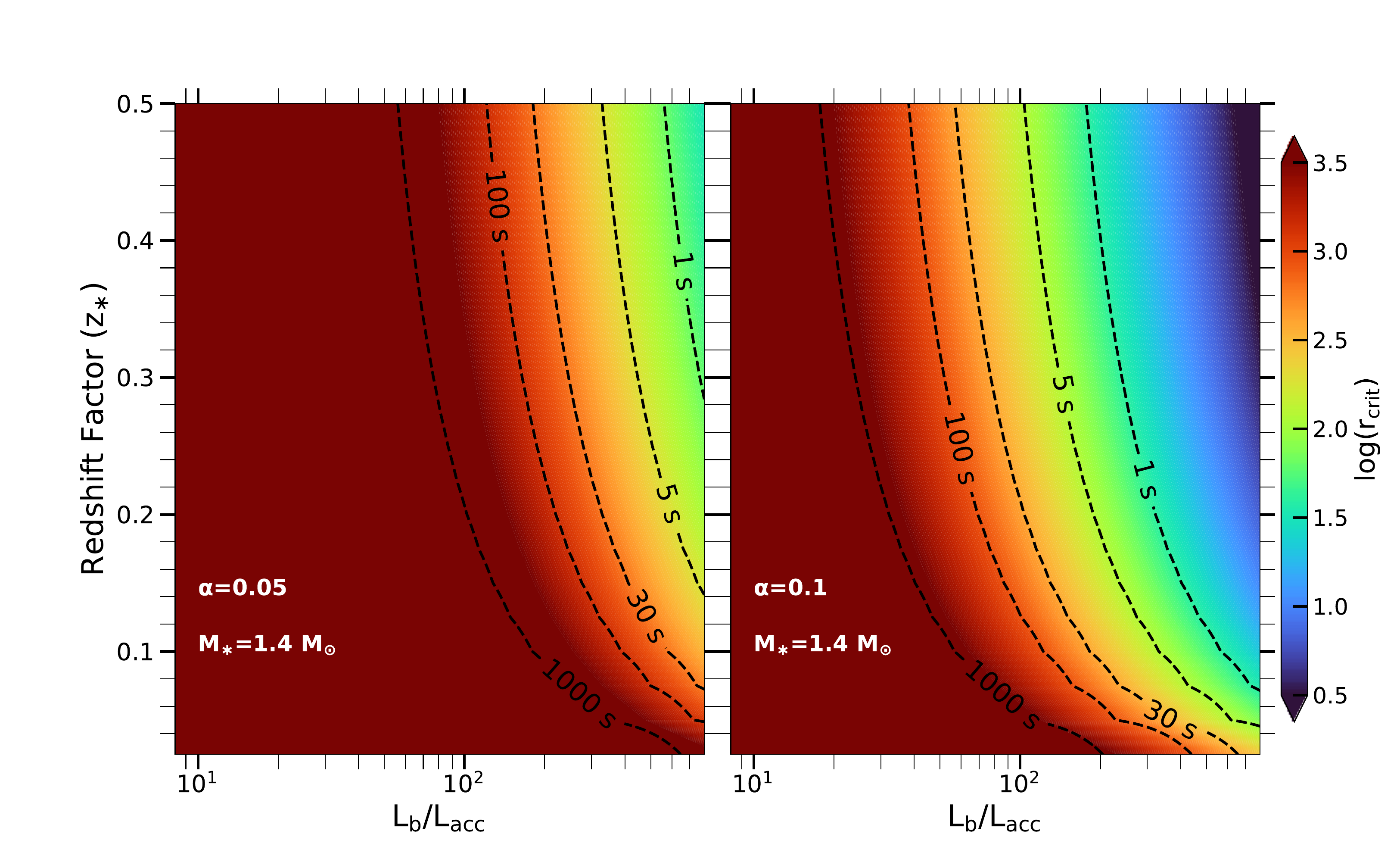}
  \caption{The colour in each panel indicates the critical radius
    $r_{\mathrm{crit}}$ at which an accretion disc becomes unstable to
  a radiation driven warp due to an X-ray burst of luminosity $L_b/L_{\mathrm{acc}}$
  from a NS with redshift factor $z_{\ast}$ (see
  Sect.~\ref{sub:rcrit}).
  The contours plot estimates of the minimum
  burst duration $t_{\mathrm{burst}}$ needed to drive the instability
  at $r_{\mathrm{crit}}$ (Sect.~\ref{sub:bursttime}) assuming a NS mass of $1.4$~M$_{\odot}$. Both $r_{\mathrm{crit}}$ and
  $t_{\mathrm{burst}}$ significantly depend on the assumed
  viscosity parameter $\alpha$ in the disc. Considering a typical
  burst luminosity of $L_b/L_{\mathrm{acc}} \approx 100$
  \citep[e.g.,][]{minbar20}, when $\alpha=0.05$ (left
  panel) a burst duration of several hundred seconds (e.g., a
  superburst) would be necessary to drive a disc warp at $r \sim 10^3$~$r_g$. However, if
  $\alpha=0.1$ (right panel) a Type I burst lasting
  $\approx 5$~s which reaches the same luminosity can drive a warp at
  $\sim 100$~$r_g$. Longer bursts, such as an IMDB, at this luminosity
  would lead to warps developing at progressively smaller radii. The
  results show that, depending on the strength of the disc viscosity,
  radiatively driven disc warps can be driven by X-ray bursts over a
  wide range of durations.} 
  \label{fig:criteria}
\end{figure*}

Using this prescription for $\eta$, Figure~\ref{fig:criteria} shows the
estimated values of $r_{\mathrm{crit}}$ and $t_{\mathrm{burst}}$ as a
function of $z_{\ast}$ and $L_b/L_{\mathrm{acc}}$. Results are calculated for
$\alpha=0.05$ (left panel) and $0.1$ (right panel). A stellar mass of
$1.4$~M$_{\odot}$ is used for estimating $t_{\mathrm{burst}}$ \citep[e.g.,][]{lattimer12}.
The panels show that both $r_{\mathrm{crit}}$ (the colour map) and
$t_{\mathrm{burst}}$ (the contours) are
strong functions of $L_b/L_{\mathrm{acc}}$ and $\alpha$, but are not sensitive to $z_{\ast}$
once $z_{\ast} \ga 0.2$. Recent measurements of the compactness of NSs
indicate $z_{\ast} \approx 0.2$--$0.3$ \citep{miller19,miller21}, indicating that the burst
luminosity and disc viscosity are the most important parameters for
the growth of a disc warp.

In general, Fig.~\ref{fig:criteria} shows that more luminous bursts lead to discs that
are unstable to warping at smaller radii and thus require a
shorter duration to grow the instability. To illustrate this, consider
the evolution of a burst in this Figure. Once the burst ignites, the luminosity quickly evolves from $L_b \sim L_{\mathrm{acc}}$ to its
peak luminosity, moving horizontally from left to right in one of the
panels of Fig.~\ref{fig:criteria}. While, in principle, the luminosity
of the burst during this transition is sufficient to drive a warp
instability at a large disc radius, the time the burst spends at these luminosities is too short for
the instability to take hold. Only when the burst reaches its peak
$L_b/L_{\mathrm{acc}}$ and holds that luminosity for a sustained period
of time will a disc warp be able to
grow. The larger the peak luminosity, the less time is needed in order
for an disc warp to develop, and this warp will grow at smaller
radii. However, the
growth of any disc warp will be arrested once the burst luminosity begins to
fall. During the tail of the burst, the luminosity moves from right to
left through the panel, and, depending on the speed of the decay, may evolve too
quickly to drive any warps at large radii along the disc. Thus, we
expect that bursts will be able to drive disc warps during the peak of
the burst, with warps growing at smaller radii during more luminous
bursts. The internal disc viscosity will work to smooth any warp
during the tail of the burst.

As the contours show the minimum burst durations, bursts that can
sustain a peak luminosity for 10s or 100s of seconds, such as a IMDB
or a superburst, will have a strong impact on the disc. In this case,
the long duration of the peak $L_b/L_{\mathrm{acc}}$ means that a
large fraction of the accretion disc could grow unstable to radiation driven
warps, and lead to substantial growth of warps at smaller radii. 

Figure~\ref{fig:criteria} shows that the viscosity of the disc
strongly impacts the critical luminosities needed to drive a warp. As
expected from the isotropic viscosity assumption, a larger value of
$\alpha$ will allow a warp to be driven at a lower luminosity. Many
accreting NSs accrete at $\sim 0.01$ of their Eddington rate
\citep{minbar20}, so $L_b/L_{\mathrm{acc}} \sim 10^2$ is a reasonable estimate for the peak
luminosity of an X-ray burst that does not exhibit
photospheric radius expansion \citep[e.g.,][]{lvt93}. Therefore, if
$\alpha=0.05$ only superbursts are predicted to produce radiation driven disc
warps at $r \sim 10^3$~$r_g$, as both IMDBs and Type I bursts are too short for a peak luminosity of $L_b/L_{\mathrm{acc}}
\sim 10^2$. However, if $\alpha = 0.1$, then Type I bursts with
peak luminosities that last $\sim 5$~s and IMDBs will both be able to
drive disc warps. Therefore, if observational evidence for disc warps can
be obtained from Type I X-ray bursts, then this would indicate that
$\alpha \ga 0.1$ in those discs. Conversely, if the effects of disc
warps are not found during superbursts, then $\alpha \la 0.05$
in those systems. 

\section{Evolution of the Disc Warping Instability During an X-ray Burst}
\label{sect:evolution}
The results of the previous section showed that, depending on the
strength of the disc viscosity, X-ray bursts may lead to a radiation
driven warp instability in the surrounding accretion disc. In
particular, X-ray
bursts with durations longer than $\sim 10$~s, such as IMDBs and
superbursts, could potentially cause a large fraction of the disc to
become unstable to warping. To understand how this instability will
impact the disc geometry, we must solve the evolution
equation for a warped accretion disc subject to the radiative torque
provided by an X-ray burst.

\subsection{Overview}
\label{sub:overview}
To describe the evolution of a warp in an Keplerian accretion disc, we first
define the local angular momentum density \citep{pringle92},
\begin{equation}
  \label{eq:L}
  \mathbf{L}(R,t)=(GM_{\ast}R)^{1/2} \Sigma \bm{\ell},
\end{equation}
where $\bm{\ell}$ is a unit vector normal to the annulus at radius
$R$ at time $t$. To define $\bm{\ell}$, consider an $xyz$ coordinate system located
at each annulus, centered on the disc and with the $x$-axis pointing to
the NS. With $\beta(R,t)$ as the local tilt angle of the disc and
$\gamma(R,t)$ as the twist angle (i.e., the azimuth of the tilt),
$\bm{\ell}$ is
\begin{equation}
  \label{eq:ell}
  \bm{\ell}(R,t)=(\cos \gamma \sin \beta, \sin \gamma \sin \beta,
  \cos \beta).
\end{equation}
Thus, $\bm{\ell}$ measures the local warp of the disc, and the evolution of
$\mathbf{L}(R,t)$ describes the development of the warp. The strength
of the warp at $R$ is measured by the local dimensionless
warp amplitude, defined as \citep[e.g.,][]{ogilvie99}
\begin{align}
  \label{eq:psi}
  \Psi & = R \left | {\partial \bm{\ell} \over \partial R} \right |
  \nonumber \\
  & = R \left ( \sin^2 \beta \left ( {\partial \gamma \over \partial
    R} \right)^2 + \left ( {\partial \beta \over \partial R} \right)^2
  \right )^{1/2}.
\end{align}

The evolution of a warped accretion disc in response to radiative
torque density $\mathbf{T}(R)$ in the viscous limit (i.e.,
$\alpha > H/R$) is \citep[e.g.,][]{pringle96,od01,martin19}
\begin{align}
  \label{eq:dLdt}
{\partial \mathbf{L} \over \partial t} & = {3 \over R} {\partial \over
  \partial R} \left [ {R^{1/2} \over \Sigma} \mathbf{L} {\partial
    \over \partial R} (\nu_1 \Sigma R^{1/2}) \right ] \nonumber \\ 
 & + {1 \over R} {\partial \over \partial R} \left [ \left ( \nu_2 R^2
  \left \lvert{\partial \bm{\ell} \over \partial R}\right
  \rvert^2 -{3 \over 2}\nu_1 \right ) \mathbf{L} \right ] \nonumber \\
 & + {1 \over R} {\partial \over \partial R} \left [ {1 \over 2} \nu_2
  R \left \lvert \mathbf{L} \right \rvert {\partial
    \bm{\ell} \over \partial R} \right ] + \mathbf{T}.
\end{align}
As in Sect.~\ref{sect:criteria}, the shear viscosities are assumed to
be isotropic, so that
\begin{equation}
  \label{eq:nus}
  \nu_1=\alpha \left( {H \over R} \right )^2
  (GM_{\ast}R)^{1/2} \hspace{0.5cm}\mathrm{and}\hspace{0.5cm} \nu_2 = {\nu_1 \over 2\alpha^2}.
\end{equation}
This assumption is valid for small warps ($\Psi < 1$) growing in the
linear regime. To a good approximation, $(H/R)$ can be considered constant for gas-pressure
dominated accretion discs far from the inner edge \citep[e.g.,][their Eq. 8]{sz94}. Thus, we fix
$(H/R)=0.0155$ for all the calculations presented in this section
which is sufficiently smaller than the two values of $\alpha$
considered ($\alpha=0.05$ and $0.1$)\footnote{Similar to
Sect.~\ref{sub:bursttime}, we neglect the increase in $H$ due to X-ray
heating during the burst. \citet{fbb19} found that $H$ approximately
doubled, which would still imply $(H/R) < \alpha$ and therefore
Eq.~\ref{eq:dLdt} remains the appropriate limit.}.

For accretion discs around compact objects, it is useful to re-write Equation~\ref{eq:dLdt} in terms of $r=R/r_g$,
$m=M_{\ast}/M_{\odot}$, and $\sigma=\Sigma/\Sigma_0$, where $\Sigma_0$
is a scaling factor for the surface density. We leave the units of time as seconds in order to more easily compare to X-ray burst
lightcurves. The evolution equation can now be written as
\begin{align}
  \label{eq:dLdt2}
{\partial \mathbfcal{L} \over \partial t} & = \left( M_{\odot}
 \over r_g^3 \right )^{1/2} \left \{ {3\alpha (H/R)^2 \over r}
{\partial \over \partial r} \left [ r^{1/2} (Gmr)^{1/2}
  \bm{\ell} {\partial \over \partial r} (\left \lvert
  \mathbfcal{L}\right \rvert r^{1/2}) \right ] \right. \nonumber \\ 
 & + {1 \over r} {\partial \over \partial r} \left [ \left ( \mu_2 r^2
  \left \lvert{\partial \bm{\ell} \over \partial r}\right
  \rvert^2 -{3 \over 2}\mu_1 \right ) \mathbfcal{L} \right ] \nonumber \\
 & + {1 \over r} {\partial \over \partial r} \left. \left [ {1 \over 2} \mu_2
  r \left \lvert \mathbfcal{L} \right \rvert {\partial
    \bm{\ell} \over \partial R} \right ] +
\mathbfcal{T} \right \},
\end{align}
where $\mu_{1,2}=\nu_{1,2}/(r_gM_{\odot})^{1/2}$,
$\mathbfcal{L}=(Gmr)^{1/2} \sigma \bm{\ell}$, and
$\mathbfcal{T}=\mathbf{T}/M_{\odot}\Sigma_0$. Given an initial $\mathbfcal{L}$ and boundary conditions at low and high
$r$, Eq.~\ref{eq:dLdt2} can be integrated explicitly in time. Once
$\mathbfcal{L}(r)$ is known at each timestep, both $\sigma(r)$ and
$\bm{\ell}(r)$ can be determined and the properties of the
warped disc can be described. 

\subsection{Numerical Method}
\label{sub:numerical}
Equation~\ref{eq:dLdt2} is integrated on a linear radial grid with
inner radius $r_{\mathrm{in}}=3$~$r_g$ and outer radius
$r_{\mathrm{out}}=401$~$r_g$ (with
$\Delta r=2$~$r_g$).
The timestep for all calculations is $\Delta
t=5\times 10^{-4}$~s, which is $\approx1.5\times$ smaller than the minimum
viscous timescale in the disc and satisfies the Courant condition.
Following \citet{martin19}, the initial surface density
profile defined on the radial grid is
\begin{equation}
  \label{eq:sigmai}
  \sigma_i(r)= \left ( {r \over r_{\mathrm{in}}} \right
  )^{-1/2} \left [ 1 - \left ( {r_{\mathrm{in}} \over r} \right
    )^{1/2} \right ] (1-e^{r-r_{\mathrm{out}}}), 
\end{equation}
which reaches a maximum at $11$~$r_g$. This profile drives the surface
density to zero at the inner and outer radii, enforcing zero torque
boundary conditions. We also ensure that $d\bm{\ell}/dr=0$ at both boundaries.

\begin{figure}
  \includegraphics[width=0.48\textwidth]{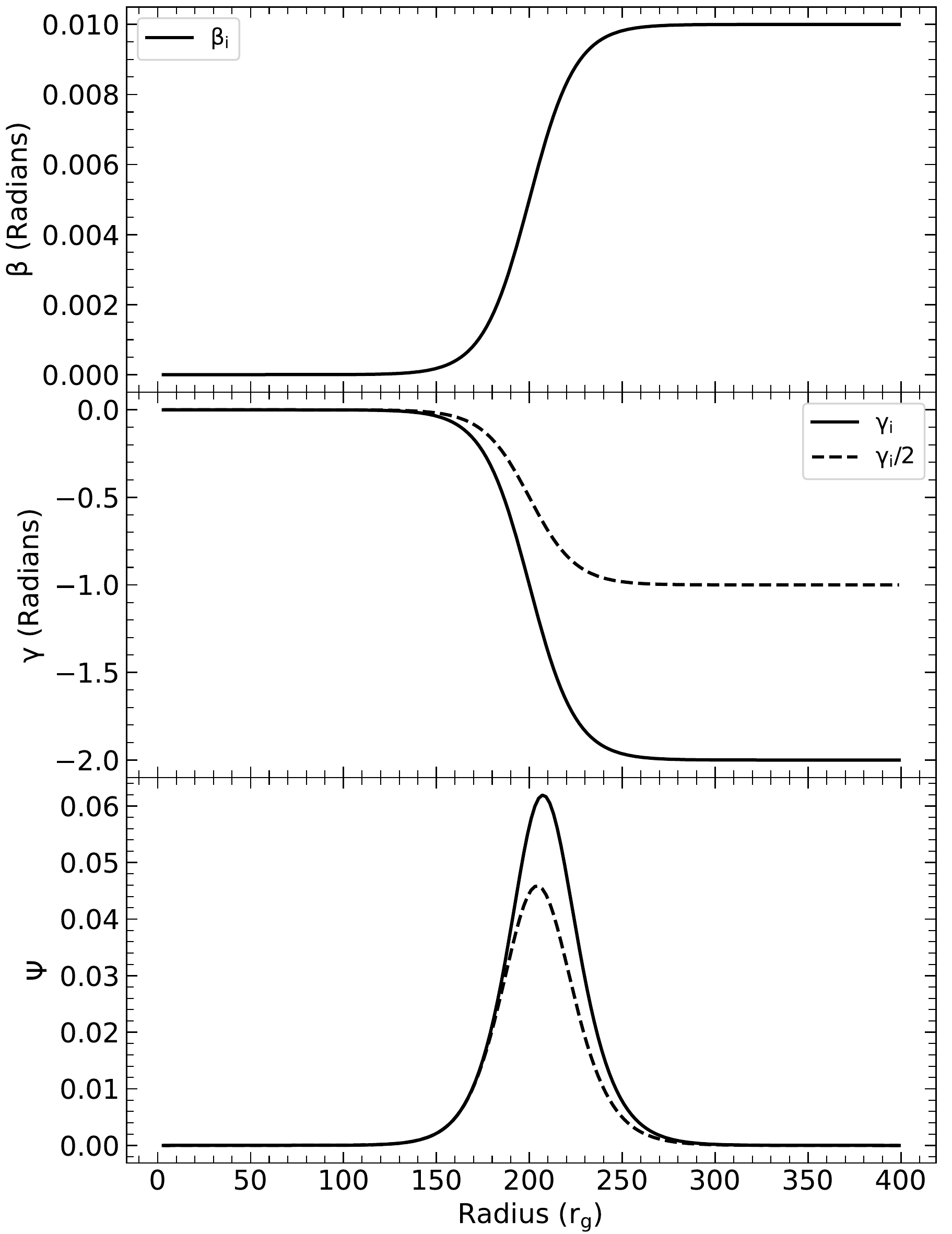}
  \caption{The top panel shows the initial disc tilt angle calculated using
    Eq.~\ref{eq:betai}. The initial twist angle $\gamma_i$
    (Eq.~\ref{eq:gammai}) is shown as the solid line in the middle
    panel, while the dashed line plots $\gamma_i/2$. The lower panel
    plots the dimensionless warp amplitude $\Psi$ (Eq.~\ref{eq:psi})
    for the two different initial warps. In both cases the disc starts
  with a small warp at the halfway point along the disc.}
    \label{fig:betainit}
\end{figure}

\citet{pringle97} found that the growth and evolution of a radiative
driven warp depends on the initial tilt and twist profiles of the
disc. Here, we define the initial tilt and twist of the disc as \citep{pringle97,martin19}
\begin{equation}
  \label{eq:betai}
  \beta_i(r)=(0.01~\mathrm{rad}) \left [ {1 \over 2} \tanh \left ( {r - r_{\mathrm{warp}} \over r_{\mathrm{width}}} \right ) + {1 \over 2} \right ]
\end{equation}
and
\begin{equation}
  \label{eq:gammai}
  \gamma_i(r)=(-2~\mathrm{rad})\left [ {1 \over 2} \tanh \left ( {r - r_{\mathrm{warp}} \over r_{\mathrm{width}}} \right ) + {1 \over 2} \right ],
\end{equation}
where $r_{\mathrm{warp}}=200$ and $r_{\mathrm{width}}=25$. These
functions describe a disc which possesses a small warp around the
midpoint of the disc (Figure~\ref{fig:betainit}). This warp could
develop between X-ray bursts from the persistent emission emanating from the
NS boundary layer and disc corona, a wind moving across the disc
surface \citep{quillen01}, or dynamical
effects from the binary companion star \citep[e.g.,][]{martin14,
  flm15}. To illustrate
the dependence on the initial warp, we will also consider models where
the starting $\gamma_i(r)$ is reduced by a factor of two (dashed line
in Fig.~\ref{fig:betainit}). 

The radiation torque density on the disc at radius $r$ due to a
luminosity $L_{\ast}$ is \citep{pringle96,od01}
\begin{equation}
  \label{eq:T}
  \mathbf{T}=-{L_\ast \over 12 \pi r c} (\bm{\ell} \times
  \mathbf{a}),
\end{equation}
where $\mathbf{a}$ is a dimensionless integral that is proportional to
$\partial \bm{\ell}/\partial r$. We follow the
procedure described in Appendix A3 of the paper by \citet{od01} to compute $\mathbf{T}$
at each timestep including the effects of self-shadowing by regions of
the disc inwards of $r$. The calculation of $\mathbfcal{T}$ for
use in Eq.~\ref{eq:dLdt} requires both $\mathbf{T}$ and the normalization of the surface
density profile, $\Sigma_0$, which is determined from
$\nu_1\Sigma=\dot{M}/3\pi$ under the assumption of a steady accretion rate. Using
the expression for $\nu_1$ from Eq.~\ref{eq:nus}, and noting that
$\Sigma \approx \Sigma_0 (r/r_{\mathrm{in}})^{-1/2}$ far from the
boundary (Eq.~\ref{eq:sigmai}), we find 
\begin{equation}
  \label{eq:sigma0}
  \Sigma_0={c \dot{M} \over 3 \pi \alpha \left ( {H \over R} \right
    )^2 GM_{\odot} m r_{\mathrm{in}}^{1/2}}
 \end{equation}
This equation is evaluated using $\dot{M}=2\times
10^{-10}$~M$_{\odot}$~yr$^{-1}$, equivalent to approximately
1\% of the Eddington rate \citep[e.g.,][]{minbar20}. The radiation torque is only applied for radii
$9\ r_g \leq r \leq 393\ r_g$ to avoid regions where the surface density $\sigma$
is falling rapidly and to remain consistent with the zero torque boundary
conditions.

To both ensure the zero-torque boundary condition and to mimic the
effect of a steady accretion rate, $\mathbfcal{L}$ is set to its
initial value for radii $r > 393$~$r_g$ at the end of every timestep. Thus,
the outer boundary maintains the small twist and tilt defined by
Eqs.~\ref{eq:betai} and~\ref{eq:gammai}. At the inner boundary, the
disc would normally terminate at a boundary layer on the surface of
the NS. However, modeling the boundary layer and its effect on the disc is beyond the scope of the
current paper, so we follow a similar procedure and set
$\mathbfcal{L}$ to its initial value for $r < 9$~$r_g$. As a result,
any disc warp is removed at very small radii, as might be expected
when interacting with a boundary layer. Appendix~\ref{app:bndry} illustrates the
results when this condition is relaxed and disc warps are allowed to
reach the inner boundary. 

\begin{figure}
  \includegraphics[width=0.5\textwidth]{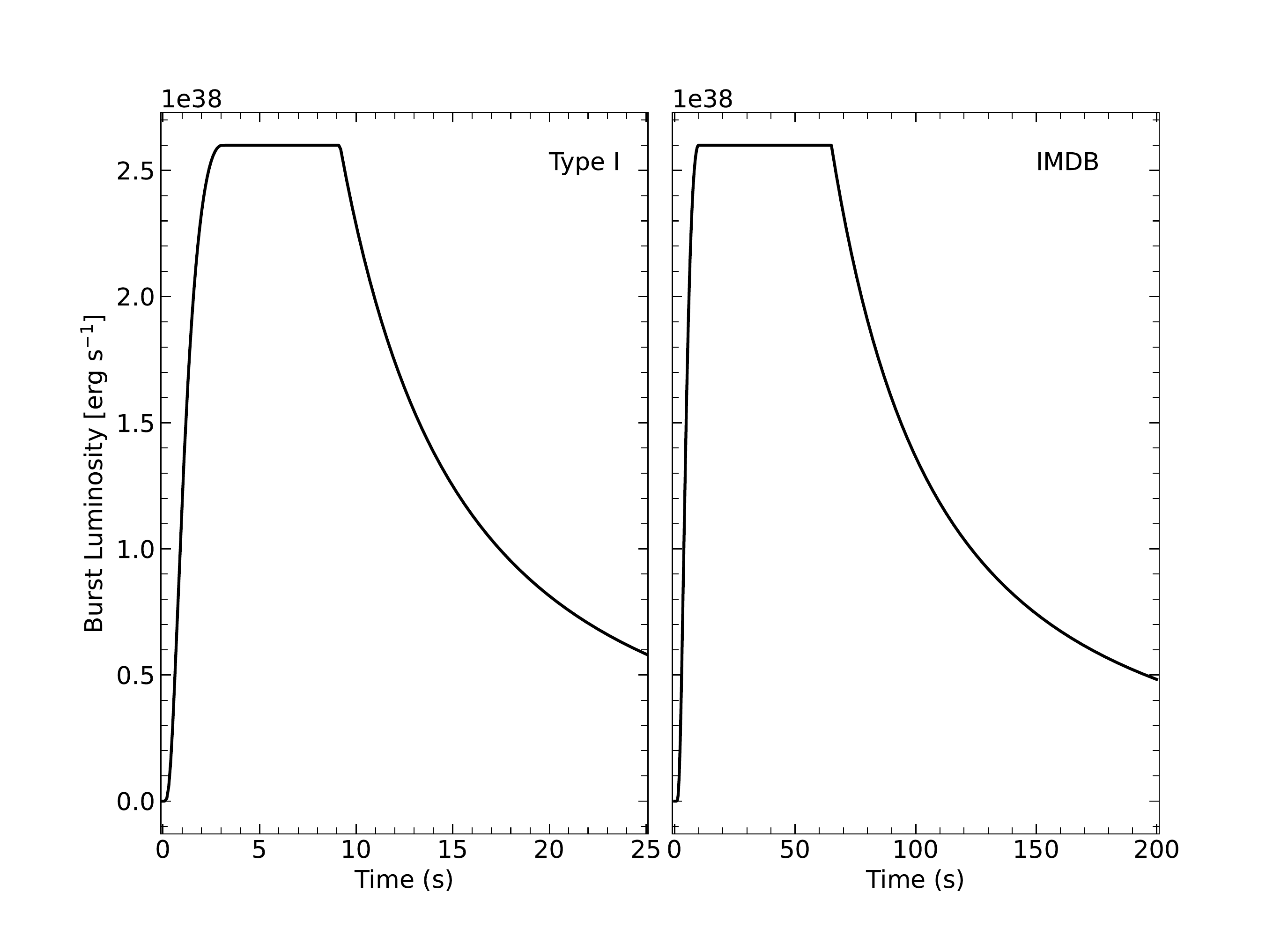}
  \caption{The radiative disc warp instability will be explored for
    two types of bursts, with lightcurves described by
    Eqs.~\ref{eq:lightcurve} and~\ref{eq:plateau}. The first is a 25~s
    long Type
    I X-ray burst (left panel) which
    peaks at $2.8$~s, followed by a $6$~s long plateau and a power-law
    decay. The right panel shows
    the light curve for a $200$~s IMDB that peaks at $10$~s and has a
    $55$~s long plateau before the power-law decay. In both cases, the
    peak luminosity of the burst is $2.6\times 10^{38}$~erg~s$^{-1}$, so that
  $L_{b}/L_{\mathrm{acc}} \approx 100$.}
    \label{fig:lightcurves}
\end{figure}

The evolution equation (Eq.~\ref{eq:dLdt2}) is integrated over time with $L_b$
following the lightcurve of either a 25~s long Type I burst or a IMDB
that lasts 200~s. The lightcurves of both bursts are described by an
exponential rise, followed by a plateau and a power-law decay
\citep[e.g.,][]{barriere15,pike21},
\begin{equation}
  \label{eq:lightcurve}
  L_b= \left \{ \begin{array}{ll}
    L_0 \exp \left (-{\tau_R \over t-t_s}-{t-t_s \over \tau_D} \right ) &
    \mbox{$0 \leq t \leq t_{\mathrm{peak}}$ }, \\
    L_{\mathrm{plateau}} & \mbox{$t_{\mathrm{peak}} < t \leq t_{\mathrm{tail}}$}, \\
    L_{\mathrm{plateau}} \left ( {t-t_s \over t_{\mathrm{tail}} - t_s}
    \right )^{-\Gamma} & \mbox{$t > t_{\mathrm{tail}}$},
  \end{array}
  \right.
\end{equation}
where $t_s$ is the burst start time, $\tau_R$ and $\tau_D$ are the characteristic rise and
decay times, $t_{\mathrm{peak}}=\sqrt{\tau_R \tau_D}$ is the time to
reach peak luminosity, and
$t_{\mathrm{tail}}=t_{\mathrm{peak}}+t_{\mathrm{plateau}}$ is the
start time
of the power-law decay, following a plateau of duration
$t_{\mathrm{plateau}}$. The plateau luminosity is given by
\begin{equation}
  \label{eq:plateau}
  L_{\mathrm{plateau}}=L_0 \exp \left (-2 \sqrt{\tau_R \over \tau_D}
  \right),
\end{equation}
where $L_0$ is a constant determined by $L_{\mathrm{plateau}}$,
$\tau_R$ and $\tau_D$. We set $L_{\mathrm{plateau}}=2.6\times
10^{38}$~erg~s$^{-1}$, which is approximately the Eddington
luminosity for a $1.4$~M$_{\odot}$ NS \citep{lvt93}, for both the Type
I burst and the IMDB models. Given the assumed value of $\dot{M}$
mentioned above, this luminosity implies $L_b/L_{\mathrm{acc}} \approx
100$. The timing parameters for the Type I
bursts are set to $\tau_R=2$~s, $\tau_D=5$~s and
$t_{\mathrm{plateau}}=6$~s \citep{pike21,guver22}, while they are
$\tau_R=10$~s, $\tau_D=10$~s and $t_{\mathrm{plateau}}=55$~s for the
IMDB models \citep[e.g.,][]{igb11}. The remaining parameters are the
same in both scenarios: $t_{\mathrm{start}}=-0.1$~s and $\Gamma=1.5$
\citep{pike21}. The lightcurves of both the Type~I burst and the IMDB
are shown in Figure~\ref{fig:lightcurves}.

The growth and evolution of the disc warp will depend on the
value of the viscosity parameter $\alpha$
(Sect.~\ref{sect:criteria}). Therefore, in addition to the dependence
on the initial disc warp, the effects of both the Type I burst and the IMDB on
the disc are studied assuming either $\alpha=0.05$ or $0.1$.
%
Our
calculation procedure saves the local disc tilt, $\beta(r)$, the local disc
twist, $\gamma(r)$, and the surface density, $\sigma(r)$, between
$r_{\mathrm{in}}$ and $r_{\mathrm{out}}$ every $0.1$~s. From
$\beta(r)$ and $\gamma(r)$, we calculate the warp amplitude $\Psi(r)$
(Eq.~\ref{eq:psi}) as a function of time, and use this quantity to
measure the evolution and strength of any disc warp that develops. Interested readers are encouraged to view
the movies found in the online Supplementary material to watch the
full time evolution of $\Psi$, $\sigma$ and $\beta$.

\subsection{Results}
\label{sub:evolres}
\subsubsection{Type I Bursts}
\label{subsub:TypeIresults}
\begin{figure*}
  \includegraphics[width=0.70\textwidth]{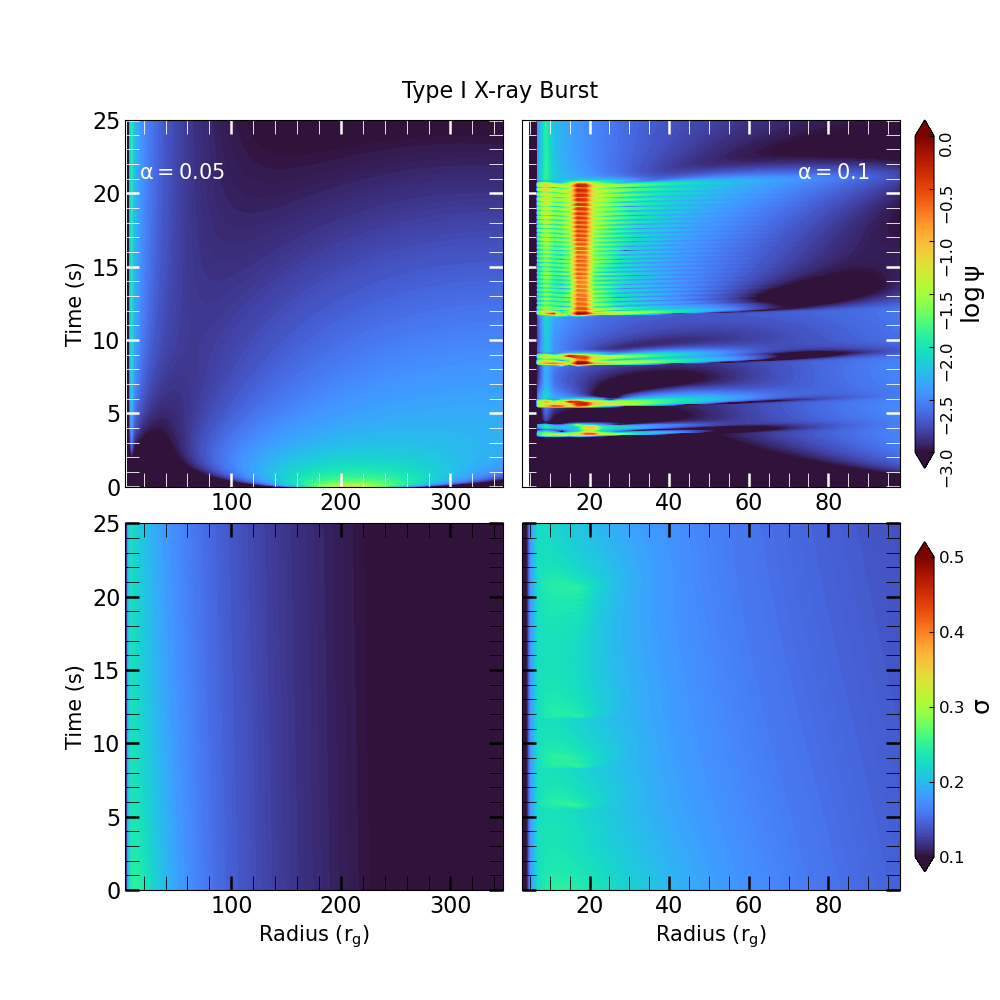}
  \caption{The top row shows space-time plots of the disc warp
    amplitude $\Psi$ (Eq.~\ref{eq:psi}) during a Type I X-ray burst with
    $\alpha=0.05$ (left) and $0.1$ (right). Space-time plots of the
    disc surface density $\sigma$ are shown in
    the bottom row. The burst does not cause a warp to grow in a disc
    with $\alpha=0.05$, consistent with the expectations from
    Fig.~\ref{fig:criteria}. In this case, the surface density slowly
    decreases due to the enhanced radial transport from the vertical
    viscosity $\nu_2$ \citep[e.g.,][]{pringle92}. Rapid growth and decay of disc warps at $r
    \la 40$~$r_g$ is found when $\alpha=0.1$, similar to the
    expectations of Fig.~\ref{fig:criteria}. The warping occurs during
  both the plateau and tail of the burst and are accompanied by an
  increase in the surface density within $\approx 20$~$r_g$. The
  `pulses' of warp during the tail of the burst are separated by
  $\approx 0.3$~s, in agreement with the vertical viscous timescale of
  the disc at these radii (Fig.~\ref{fig:timescales})}.
    \label{fig:type1}
\end{figure*}

Figure~\ref{fig:type1} displays space-time plots of the disc warp
amplitude $\Psi$ (Eq.~\ref{eq:psi}; top row) and surface density
$\sigma$ (bottom row) found from integrating the
evolution equation (Eq.~\ref{eq:dLdt2}) during a Type I burst. The
initial disc warp is described by $\beta_i$ and $\gamma_i$
(Fig.~\ref{fig:betainit}), and the initial surface density is defined
by Eq.~\ref{eq:sigmai}. According to Sect.~\ref{sub:res1}, a Type I
burst with a peak luminosity of $L_b/L_{\mathrm{acc}} \approx 100$ is
not expected to drive a disc warp when $\alpha=0.05$, and the
left-hand side of Figure~\ref{fig:type1} confirms this prediction.
Close inspection of the $\Psi$ panel for the $\alpha=0.05$ calculation
shows that the initial warp centered at $200$~$r_g$ dissipates and
propagates to higher and lower radii during the first $5$ seconds of the
burst, with no significant warp growth occurring during any part of the
burst. Similarly, the disc surface density panel only shows a
steady reduction of $\sigma$ in the inner disc due to the enhanced
radial transport caused by the vertical viscosity $\nu_2$. The lack of
any disc warping during the $\alpha=0.05$ Type I burst calculation
supports the validity of the analytical estimates of
Sect.~\ref{sub:res1}, and also indicates that our numerical method is
not introducing spurious warps into the calculations.

Qualitatively different behavior is observed from the $\alpha=0.1$
calculation (right-hand side of Fig.~\ref{fig:type1}). In this scenario,
substantial and repeated disc warps develop and dissipate at $r \la
40$~$r_g$ starting at $\approx 3.5$~s and lasting through both the
plateau and, in particular, the tail of the X-ray burst. The growth of
disc warps in the $\alpha=0.1$ Type I burst calculation is consistent
with the predictions of Fig~\ref{fig:criteria}, although the expected
critical radius is predicted to be $\sim 100$~$r_g$. The lower-right panel of
Fig.~\ref{fig:type1} shows that these warp events are accompanied by
increases in the disc surface density, resulting from the temporary
build-up of material over the small range of radii in the warp.

\begin{figure*}
  \includegraphics[width=0.80\textwidth]{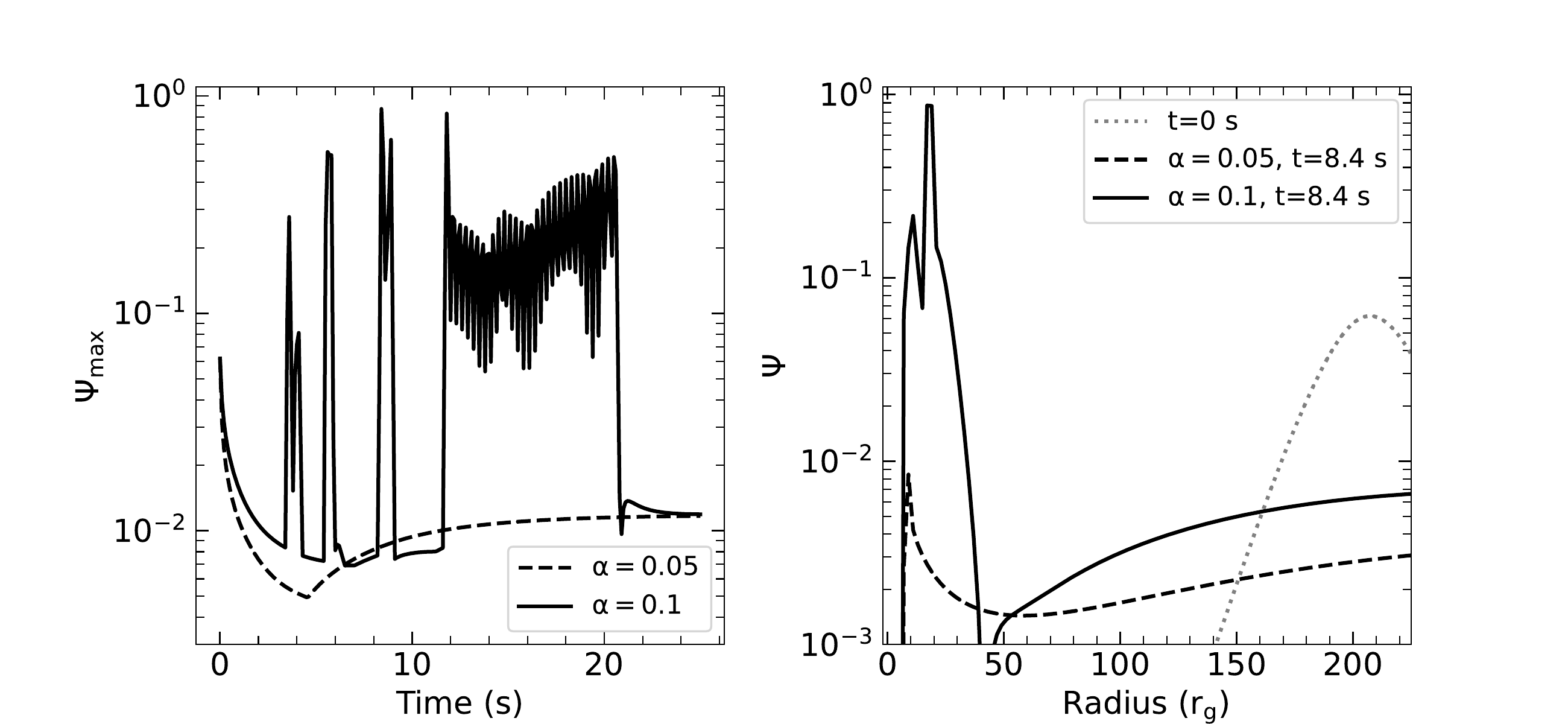}
  \caption{(Left) The time evolution of the maximum value of the disc
    warp amplitude $\Psi$ during the Type I burst calculations with
    $\alpha=0.05$ (dashed line) and $0.1$ (solid line). The initial
    disc warp is quickly reduced by the vertical viscosity in the
    $\alpha=0.05$ model and reaches a steady state determined by the
    boundary conditions. As seen in Fig.~\ref{fig:type1}, disc warps
    grow and decay rapidly in the $\alpha=0.1$ calculation. The
    maximum value of $\Psi=0.87$ occurs at $t=8.4$~s and
    $r=17$~$r_g$ when the burst luminosity is $2.6\times
    10^{38}$~erg~s$^{-1}$. The rapid pulsations of warp in the tail of
  the burst all have $\Psi_{\mathrm{max}} \la 0.5$. These warps
  therefore remain in the linear regime. (Right) The radial profiles
  of $\Psi$ when $t=0$ (dotted line) and $t=8.4$~s for the
  $\alpha=0.05$ (dashed line) and $0.1$ (solid line). At this time in
  the $\alpha=0.05$ calculation, the initial disc warp amplitude,
  which was concentrated at $r \approx 200$~$r_g$ has been reduced and
  propagated across the disc. The $\alpha=0.1$ calculation exhibits a
  substantial, but temporary warp inwards of $\approx 40$~$r_g$. }
  \label{fig:type1psi}
\end{figure*}

Three significant disc warp events occur during the plateau phase of
the burst, each getting progressively stronger; however, $\Psi < 1$ in
all cases, indicating that the warps remain in the linear regime
(Figure~\ref{fig:type1psi}; left panel). 
The
growth and decay of the warps are due to the complex interactions
between the disc viscosity attempting to flatten the disc, the
radiation torque trying to warp the disc, and the effects of
shadowing, which can significantly alter the pattern of radiation
torque on the disc \citep[see also][]{pringle97}. This is illustrated
in the right panel of Fig.~\ref{fig:type1psi} which plots $\Psi(r)$ at
the time that $\Psi$ reaches its maximum (at $8.4$~s). At this point,
$\Psi$ peaks at $0.87$ at $r=17$~$r_g$. The warp is concentrated in
the inner regions of the disc, where the flux and radiative torque is
largest, but these radii also have the shortest viscous and diffusion
timescales (see below), which will act to quickly reduce the warp. The
disc just behind the warp (at slightly larger radii) is shadowed by
the warped disc inside of it, which reduces the growth of the warp
until the radiation torque can
once again dominate. 

\begin{figure*}
  \includegraphics[width=0.80\textwidth]{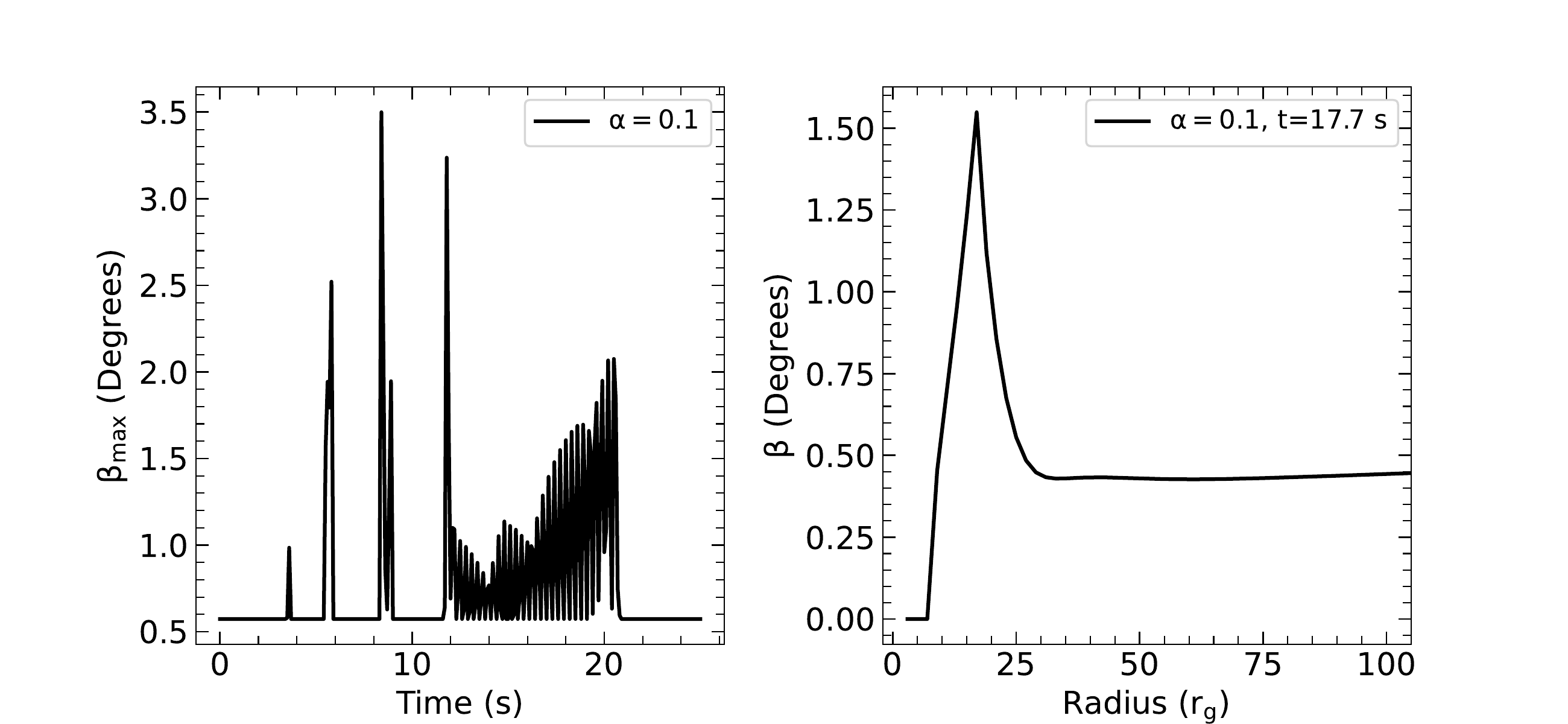}
  \caption{(Left) The time evolution of the maximum value of the disc
    tilt $\beta$ during the Type I burst calculations with
    $\alpha=0.1$ (dashed line). The maximum disc tilt of
    $\beta=3.5$~degrees is reached at $8.4$~s and a radius of
    $17$~$r_g$. The maximum $\beta$ barely exceeds $2$~degrees during
    the pulsations in the burst tail. (Right) The radial profile of
    $\beta$ across the inner quarter of the disc at $t=17.7$~s from the
  $\alpha=0.1$ calculation. The maximum tilt occurs at $17$~$r_g$.}
  \label{fig:type1beta}
\end{figure*}

The most striking aspect of the $\alpha=0.1$ warping calculation are
the repeated 'pulses' of warp during the tail of the burst that start
at $\approx 12$~s and end at $\approx 21$~s. Each of these warps have
$\Psi \la 0.5$, and peak at $17$~$r_g$. To get a more physical view of
the nature of these warps, the left panel of
Figure~\ref{fig:type1beta} plots the time evolution of maximum disc
tilt $\beta$ along the disc in degrees. 
The panel shows that the disc tilts remain small, as expected by the
linear growth approximations made in our calculation. The right 
side of Fig.~\ref{fig:type1beta} shows part of the radial profile of
$\beta(r)$ at the peak of one of the pulses during the tail of the
burst and shows that it is concentrated at radii $\la
25$~$r_g$. Therefore, while small, these warps may still subtend a
non-negligible solid angle as seen from the neutron star.

\begin{figure}
  \includegraphics[width=0.48\textwidth]{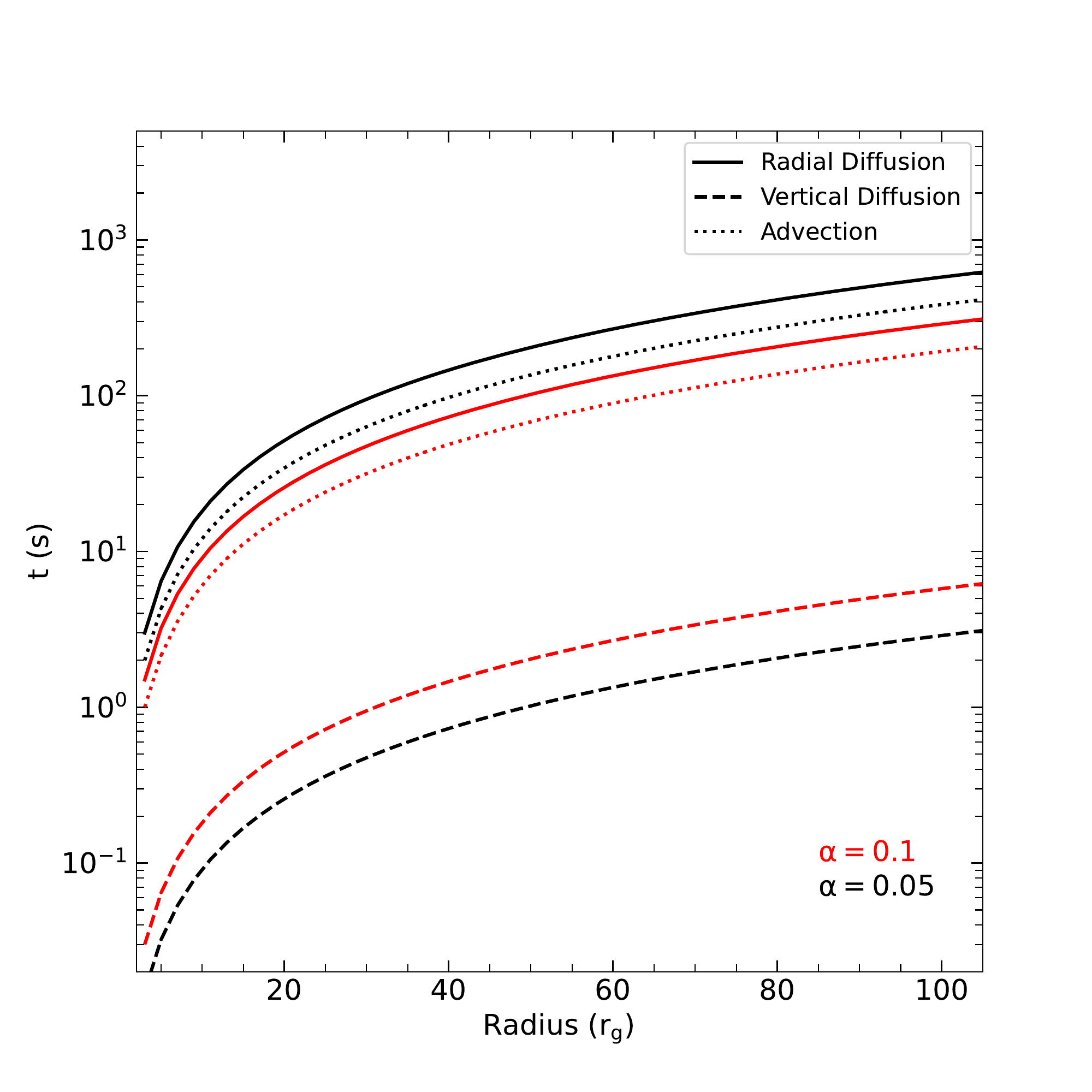}
  \caption{The radial profiles of the three different timescale
    relevant to a warped accretion disc: radial diffusion
    (Eq.~\ref{eq:tdiff}, solid
    lines), vertical diffusion (Eq.~\ref{eq:tdiff}, dashed lines), and advection (at
    $t=0$; Eq.~\ref{eq:tadv}, dotted
    lines). Timescales for calculations with $\alpha=0.1$ are shown
    in black while red lines indicate the results for $\alpha=0.05$.}
  \label{fig:timescales}
\end{figure}

The pulses of warp during the burst tail of the $\alpha=0.1$
calculation occur at regular intervals of $\approx
0.3$~s. Figure~\ref{fig:timescales} plots estimates of the three
timescales important in warped discs\footnote{The dynamical time is
$\sim 10^{-3}$~s at these radii and is off the scale of the plot.}.
The solid and dashed lines show the radial and vertical diffusion times
\begin{equation}
  \label{eq:tdiff}
  t_{\mathrm{rad,vert}} \sim {R^2 \over \nu_{\mathrm{1,2}}} \sim
  {r_g^{3/2} \over M_{\odot}^{1/2}} \left ( {r^2 \over
    \mu_{\mathrm{1,2}}} \right ),
\end{equation}
and the dotted lines plot the advection timescale
\begin{equation}
t_{\mathrm{adv}} \sim {R \over v} \sim {r \over v} \left ( {r_g^3 \over
  M_{\odot}} \right )^{1/2},
\label{eq:tadv}
\end{equation}
where the advection velocity is
\begin{equation}
  v = {1 \over r} \left ( {3 \over 2} \mu_1  - \mu_2 \Psi^2 \right ).
  \label{eq:vadv}
\end{equation}
The advection velocity is computed using the initial value of $\Psi$
which is $0$ over this range of radii
(Fig.~\ref{fig:betainit}). Therefore, $t_{\mathrm{adv}}$ in
Fig.~\ref{fig:timescales} will be a lower limit. Given both the
timescale ($\sim 0.3$~s) and radii ($\sim 20$~$r_g$) of the
pulsations, we see that these properties are consistent with the
vertical diffusion timescale in an $\alpha=0.1$ disc. This is the
primary timescale at which the disc attempts to flatten any warps. It
appears that during the tail of the Type I burst, a mode in the disc
is excited in which warp growth and destruction are locked on the same
timescale. This lasts for several seconds before the decline in burst
luminosity (and therefore radiative torque) reduces the ability of
warps to grow at the required rate.

\subsubsection{IMDBs}
\label{subsub:IMDBresults}
\begin{figure*}
  \includegraphics[width=0.70\textwidth]{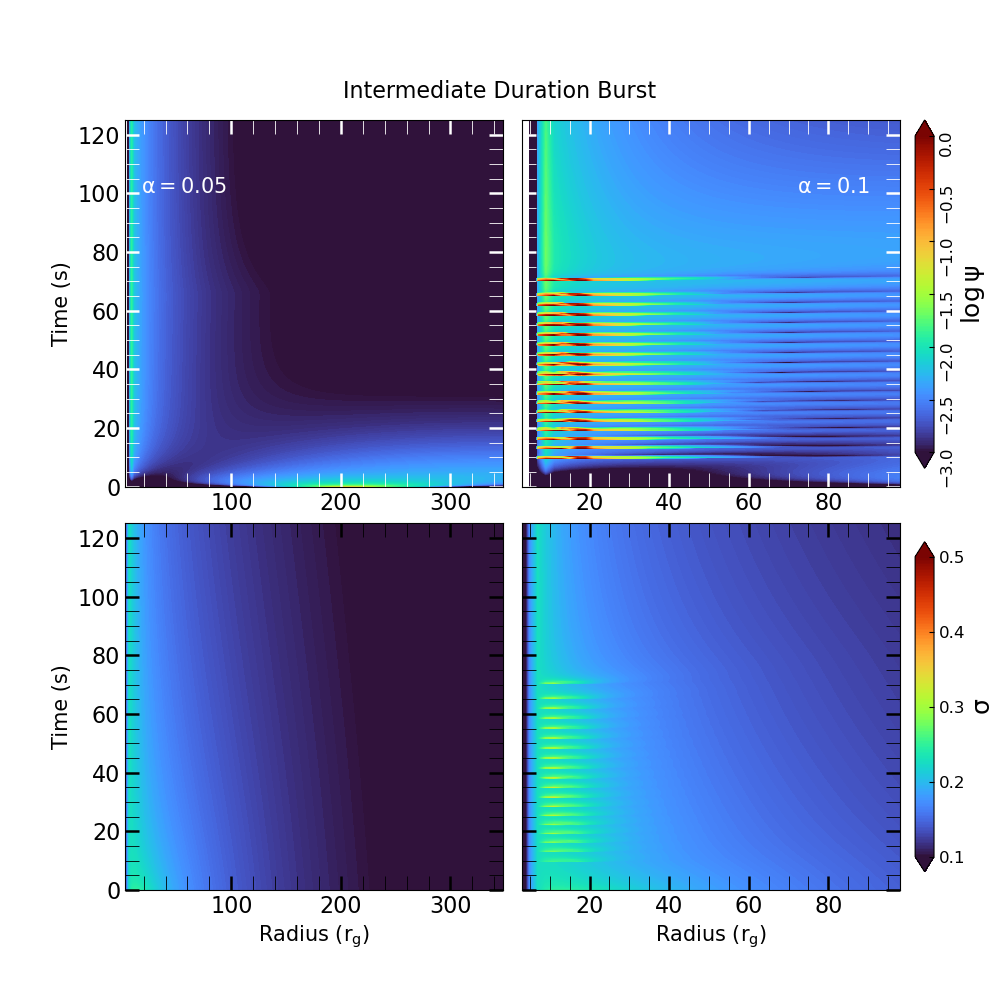}
  \caption{As in Fig.~\ref{fig:type1}, but now for the longer
    IMDB. Consistent with the analytical expectations
    (Fig.~\ref{fig:criteria}) and similar to the Type I burst, the
    IMDB does not cause a warp to grow in a disc with
    $\alpha=0.05$. When $\alpha=0.1$, roughly periodic growth and
    decay of warps occur at $r \la 50$~$r_g$ during the plateau of the
    IMDB. The pulses of warp occur at intervals of $\approx 3$--$4$~s and can exceed
    $\Psi = 1$, indicating they are entering the non-linear
    regime. The longer duration between pulses is more consistent with
    the radial diffusion timescale (Fig.~\ref{fig:timescales}). The warping is concentrated during the high-luminosity
    plateau of the burst with only one significant pulse occurring
    during the tail of the IMDB (at
    $70.4$~s).} 
    \label{fig:imdb}
\end{figure*}

The results of the warp evolution calculations for the longer IMDBs
are shown in Figure~\ref{fig:imdb}. As with the Type I burst models
discussed above, the initial disc warp is defined with $\beta_i$ and
$\gamma_i$ (Fig.~\ref{fig:betainit}). 
The left-hand side of the figure shows that no disc warp develops when
$\alpha=0.05$, similar to the Type I burst case. The results of
Sect.~\ref{sect:criteria} indicate that even longer bursts (such as
superbursts), or luminosities several times larger, would be needed to
drive a disc warp when $\alpha=0.05$.

\begin{figure*}
  \includegraphics[width=0.80\textwidth]{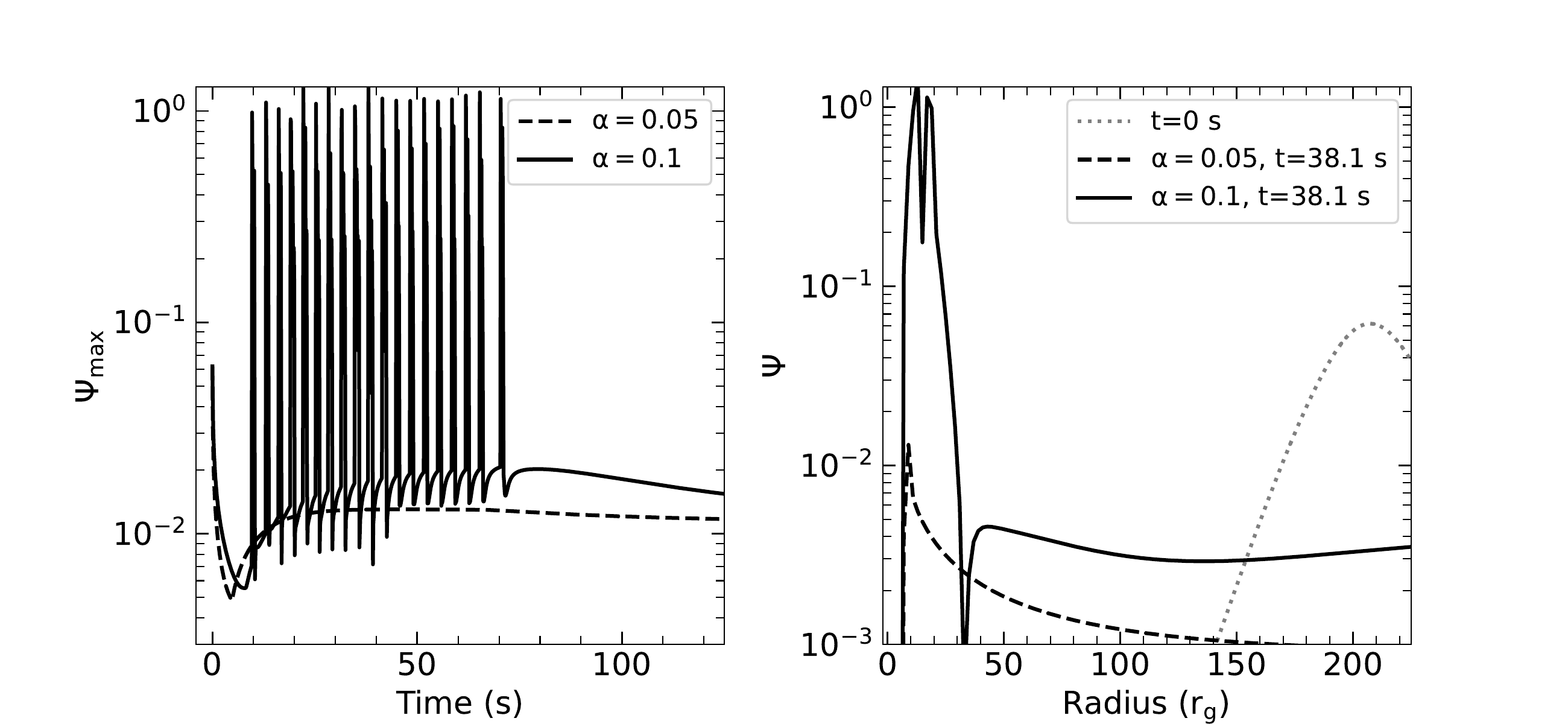}
  \caption{As in Fig.~\ref{fig:type1psi}, but now for the IMDB
    calculations. The maximum disc warp amplitude ($\Psi=1.45$) in the
    $\alpha=0.1$ model occurs at $t=38.1$~s and $r=13$~$r_g$ during
    the plateau of the IMDB. As in the Type I burst, the warps peak at
  small radii, within $\sim 20$~$r_g$.}
  \label{fig:imdbpsi}
\end{figure*}

\begin{figure*}
  \includegraphics[width=0.80\textwidth]{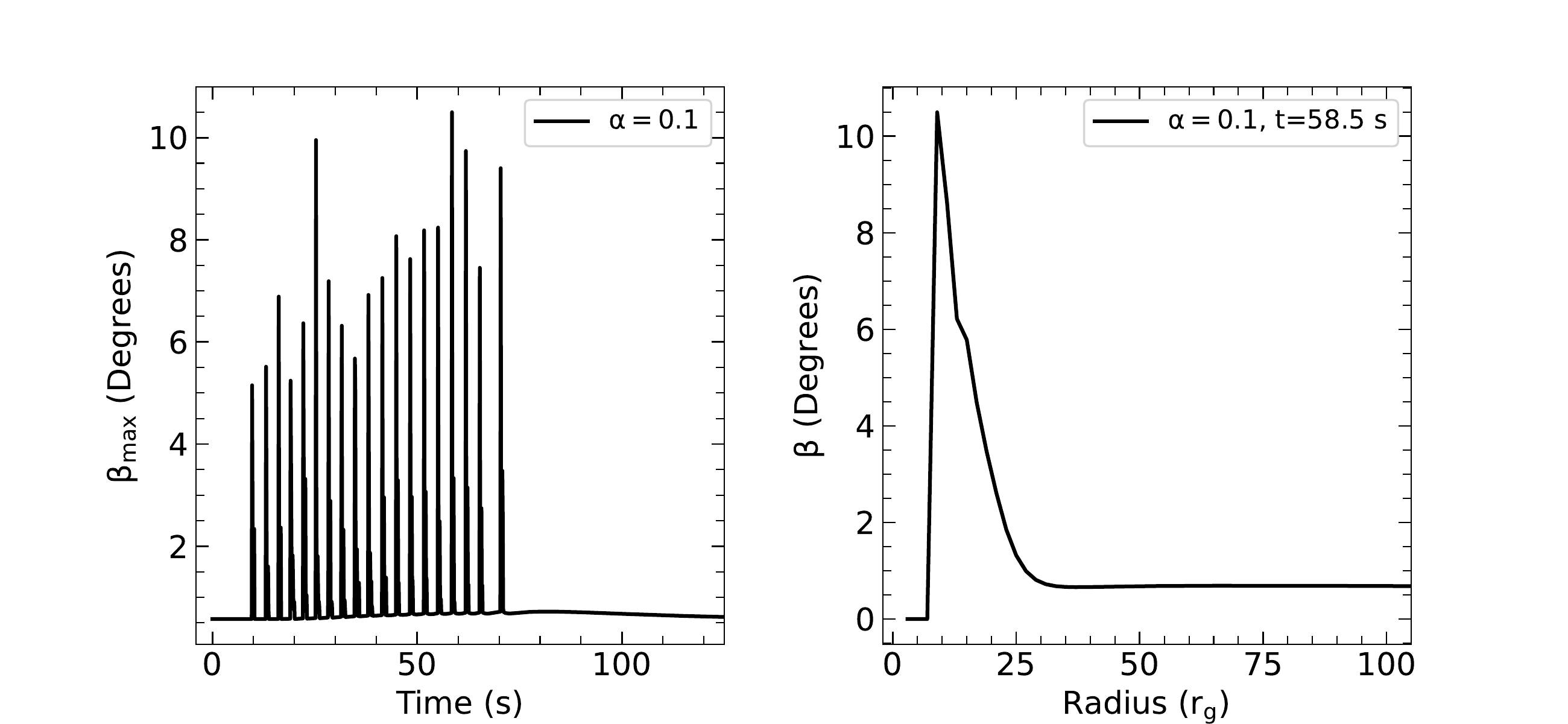}
  \caption{As in Fig.~\ref{fig:type1beta}, but now for the longer
    IMDB. The maximum disc tilt during the $\alpha=0.1$ calculation is $10.5$~degrees at $t=58.5$~s into
    the burst. The warp is concentrated at the innermost regions of
    the disc, peaking at $r=9$~$r_g$.}
  \label{fig:imdbbeta}
\end{figure*}

Turning to the $\alpha=0.1$ calculation (right side of
Fig.~\ref{fig:imdb}), we find repeated growth and decay
of warps occuring at $r \la 50$~$r_g$ during the plateau of peak luminosity in the
IMDB. The warps also lead to temporary increases in the surface
density over the same range of radii. Additional details of the warps
are shown in Figure~\ref{fig:imdbpsi}, which plots the evolution of
the peak $\Psi$ during the IMDB.
Unlike the pulsations seen in the Type I calculation, the warps found
during the IMDB sometimes reach $\Psi > 1$, with the maximum value of
$1.45$ occurring at $38.1$~s. These warps, therefore, enter the
non-linear regime, moving outside the range of validity for this
calculation. The right panel of Fig.~\ref{fig:imdbpsi} show that the
peak of the warps are concentrated within $\sim 20$~$r_g$. The larger
values of $\Psi$ naturally lead to increased disc tilts $\beta$, as
shown in Figure~\ref{fig:imdbbeta}. The largest disc tilt is
$10.5$~degrees and occurs $58.5$~s into the IMDB. 

Unlike the pulses seen in the Type I burst calculation, the repeated
disc warps in the IMDB model occur almost exclusively during the
plateau phase of the burst, with a single, final warp occurring $5.4$~s
into the tail, at $t=70.4$~s. The pulses of warp in the IMDB
calculation are separated by $\approx 3$--$4$~s, ten times longer than
the ones found in the Type I burst. This timescale is consistent with
the radial diffusion timescale at $r \la 20$~$r_g$
(Fig.~\ref{fig:timescales}). Therefore, while the warps are quickly
flattened by the vertical viscosity, the re-generation timescale is
set by how quickly the disc can bring material through to the inner
regions of the disc.

\subsubsection{Reducing the Initial Warp}
\label{subsub:halfgammai}
\begin{figure*}
  \includegraphics[width=0.70\textwidth]{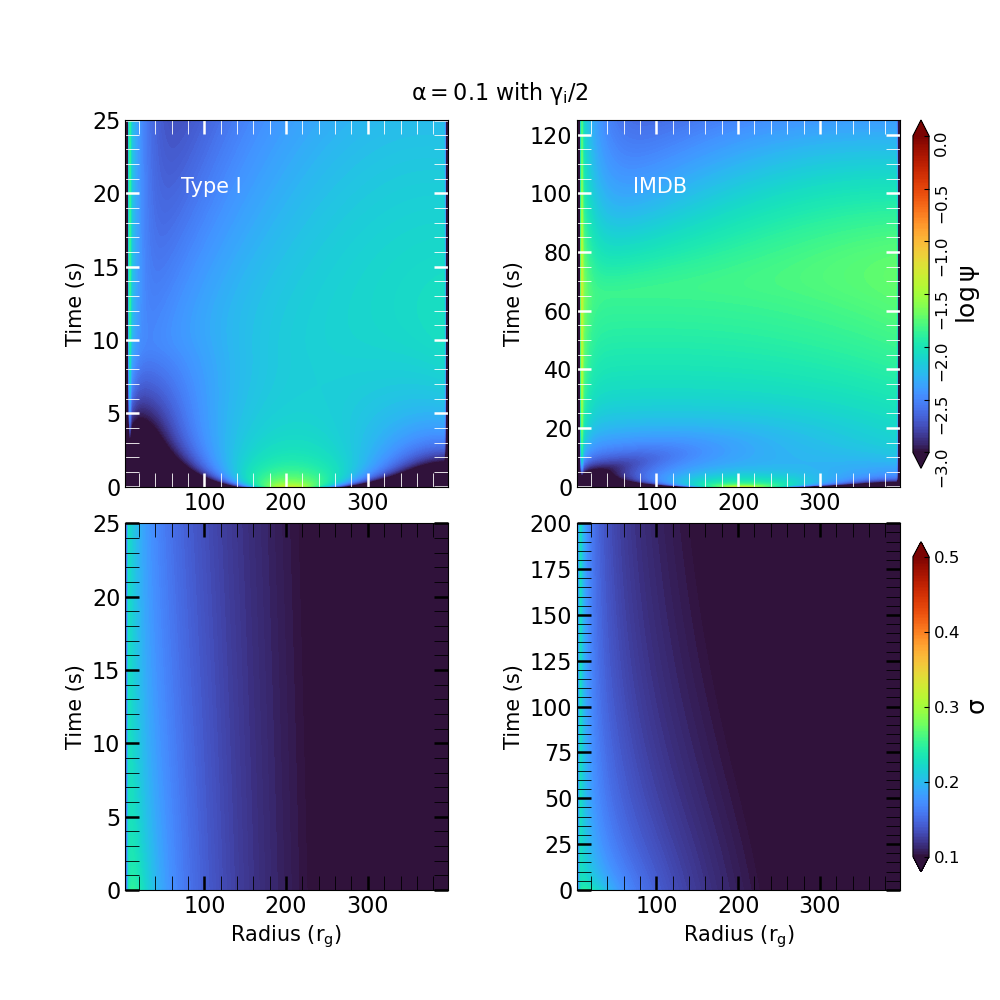}
  \caption{As in Figs.~\ref{fig:type1} and~\ref{fig:imdb}, but now
    showing the $\alpha=0.1$ results when the initial disc twist is
    $\gamma_i/2$ (see Fig.~\ref{fig:betainit}). The initial disc tilt
    is unchanged and equal to $\beta_i$. For both types of bursts, this simple change in the starting
    conditions leads to a qualitatively different evolution of disc
    warp. The pulses of warp concentrated at $r \la 50$~$r_g$ are now absent and are
    replaced with a slower growing warp spread over the entire
    disc. The warp amplitude increases steadily during the plateau of
    both bursts, before declining during the burst tail. With these
    conditions, a burst lasting $\sim 1000$~s (such as a superburst)
    would drive a significant warp over a large fraction of the disc.} 
    \label{fig:halfgamma}
\end{figure*}

As demonstrated by \citet{pringle97}, the evolution of a disc warp is
sensitive to the initial warp defined at $t=0$. We investigate this
here by repeating the $\alpha=0.1$ calculations for both the Type I
burst and IMDB with $\gamma_i/2$. The initial disc tilt remains
$\beta_i$, but the change in the starting disc twist reduces the
initial $\Psi$ of the disc (Fig.~\ref{fig:betainit}). The resulting
space-time plots of $\Psi$ and $\sigma$ are shown in
Figure~\ref{fig:halfgamma}. 
The plots show a very different warp evolution than the previous calculations
with $\gamma_i$ (Figs.~\ref{fig:type1} and~\ref{fig:imdb}). Rather
than pulsations of disc warp that rapidly form and dissipate within
$\sim 50$~$r_g$, we instead see a slowly growing disc warp that covers
a large fraction of the disc. The warp grows during the plateau phase
of each burst before declining in the tail. As a result, the warp
reaches a higher value during the IMDB, but remains safely in the
linear regime (the maximum $\Psi$ is $0.044$). Similarly, unlike the
strong warp pulsations seen earlier, the $\gamma_i/2$ warps do not result in
an increase in surface density. Interestingly, the
radii of warp growth in these $\gamma_i/2$ calculations are
closer to the values of the critical radii predicted by
Fig.~\ref{fig:criteria}, indicating that the smaller initial $\Psi$
may be important in comparing to the analytical estimates of
$r_{\mathrm{crit}}$.

\begin{figure*}
  \includegraphics[width=0.80\textwidth]{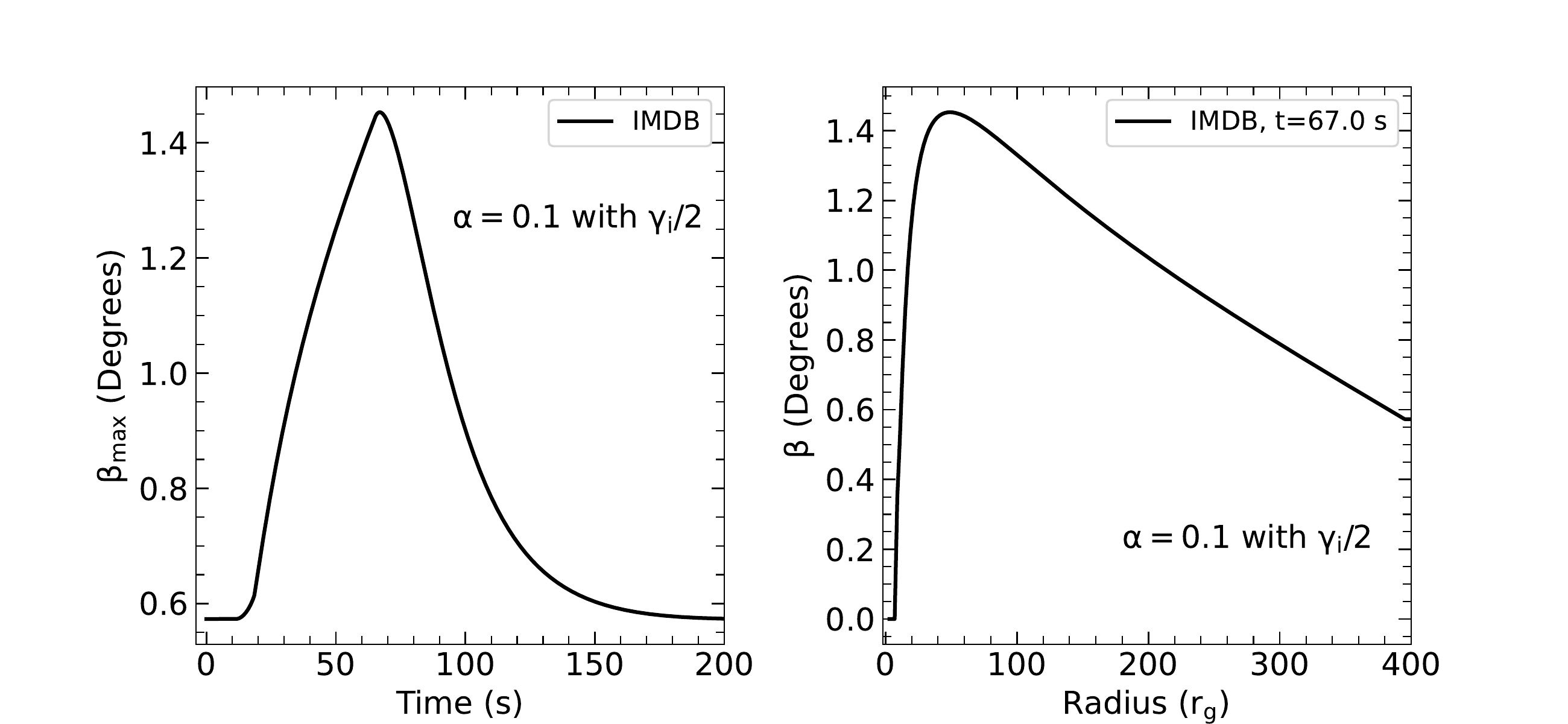}
  \caption{As in Fig.~\ref{fig:type1beta}, but now showing the results
    for the IMDB calculation with the smaller $\gamma_i/2$ starting
    disc twist. The maximum disc tilt is $1.45$~degrees which is
    reached at $67$~s and $r=49$~$r_g$. The luminosity of the burst
    has just started its decline with the luminosity at this point
    equal to $1.1\times 10^{38}$~erg~s$^{-1}$. The disc tilt increases
  steadily during the plateau phase of the burst.}
  \label{fig:halfgammabeta}
\end{figure*}

The evolution of the maximum disc tilt in the IMDB calculation with
$\gamma_i/2$ is shown in the left-hand panel of
Figure~\ref{fig:halfgammabeta}. The steady increase of
$\beta_{\mathrm{max}}$ during the plateau phase of the IMDB is clearly
apparent, with $\beta$ reaching a maximum of $1.45$~degrees at
$t=67$~s, shortly after the burst luminosity begins its decline.
The disc tilt therefore grew $\sim 1.4$~degrees in $\sim
55$~s. Assuming the same rate of growth, a superburst with a plateau
of $\sim 1000$~s (e.g., 4U~1636-53; \citealt{keek14a,keek14b}) would
generate a maximum disc tilt of $\sim 25$~degrees. The radial profile
of the warp at $t=67$~s is shown in the right panel of
Fig.~\ref{fig:halfgammabeta}. The peak tilt occurs at $r=49$~$r_g$
with a slow decline to larger radii. The profile is far less 'peaky'
than the ones found with the $\gamma_i$ starting condition (e.g.,
Fig.~\ref{fig:imdbbeta}), indicating that the complex feedback effects
from shadowing are less important in the evolution of this warp.

\begin{figure}
  \includegraphics[width=0.48\textwidth]{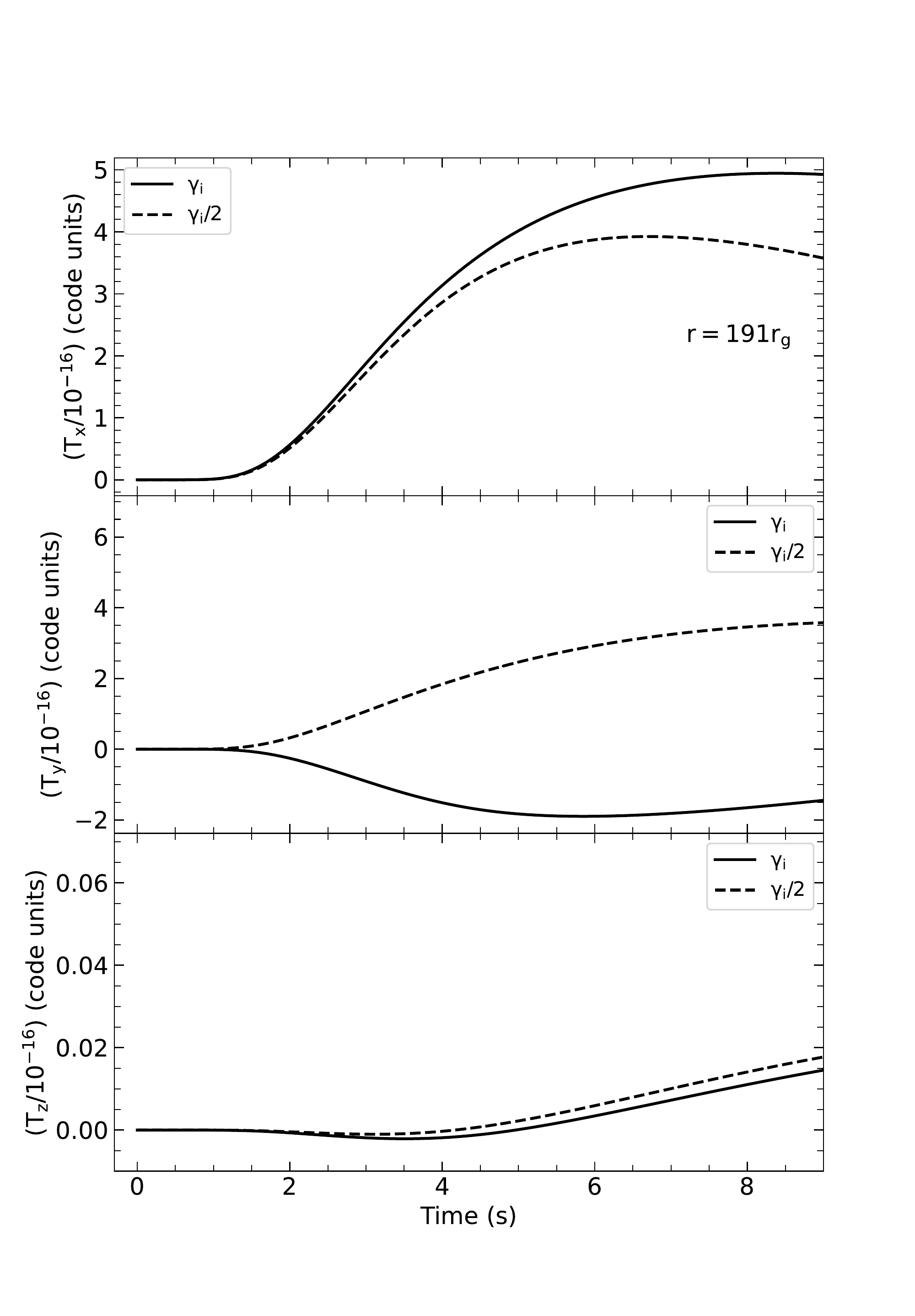}
  \caption{The components of the radiative torque density vector
    $\mathbfcal{T}$ at $r=191$~$r_g$ during the first $9$~s of the
    IMDB, $\alpha=0.1$ calculation. The solid line shows the torque
    density for the $\gamma_i$ initial disc twist. This is the
    calculation that leads to warp pulsations during the burst plateau
  (Sect.~\ref{subsub:IMDBresults}). The dashed line plots the torque
    density when the initial disc twist is $\gamma_i/2$ which results in
  a more slowly growing warp throughout the disc.}
  \label{fig:torque}
\end{figure}

To understand how a modest factor of $2$ change in $\gamma_i$ can
yield such a different outcome, Figure~\ref{fig:torque} plots the
components of the radiative torque density vector $\mathbfcal{T}$ at
$r=191$~$r_g$ during the initial $9$~s of the IMDB. 
The torque densities for the calculation starting with a disc twist
$\gamma_i$ are shown with the solid lines, while the dashed lines
plots the components of $\mathbfcal{T}$ for the $\gamma_i/2$
model. The figure show that the two torque densities at this location
(which is close to the peak of the initial $\Psi$;
Fig.~\ref{fig:betainit}) coincide for the first $\approx 1.5$~s of the
calculation before diverging. The strongest torques are in the $x$ and
$y$ directions and lead to procession of a tilted orbiting ring
\citep[e.g.,][]{pringle96}. The torques in these 2 directions show
different behaviors depending on the initial disc twist, in particular
in the $y$-direction where the torques have different signs. In the limit
of no shadowing $\mathbfcal{T} \propto \bm{\ell} \times \partial
\bm{\ell}/\partial r$ \citep{od01}, which leads to terms proportional
to $\partial \beta/\partial r$ and $\partial \gamma/\partial r$. Lowering the
initial disc twist to $\gamma_i/2$ also reduces $\partial
\gamma/\partial r$, which ultimately causes the change in sign for the
torque density in the $y$-direction. As the torque depends on the
gradients in $\gamma$ and $\beta$, these small differences will
quickly accumulate and feedback on each other, leading to the
strikingly different warp evolutions seen in the two calculations.

\section{Discussion}
\label{sect:discuss}
\subsection{Observational Implications of Burst Driven Warps}
\label{sub:observations}
The high luminosities of X-ray bursts are necessary for generating
warps at moderate disc radii during the burst. In addition, this
luminosity must be sustained for several seconds or minutes
in order to drive a warp (Fig.~\ref{fig:criteria}). Therefore, we
expect the highest probability for observing signatures of disc warps
in IMDBs or superbursts. However, we also find that the growth of a
warp significantly depends on the strength of the viscosity in the
disc. Assuming isotropic viscosity, neither Type I bursts nor IMDBs
were able to generate a disc warp when $\alpha=0.05$, while both situations
yielded warps when $\alpha$ was doubled (see
Figs.~\ref{fig:type1} and~\ref{fig:imdb}). Numerical simulations of
accretion discs with viscosities caused by the magneto-rotational
instability \citep{bhMRI91,bh98} find that the effective $\alpha$ is
both $< 0.05$ and varies with both
position and time in the flow \citep[e.g.,][]{davis10,bodo14,ryan17}. These results imply that
radiatively driven warps during bursts may be rare and only occur when
the exact combination of viscosity, luminosity and duration is
obtained at a particular location in the disc. Alternatively,
observations of accretion outbursts in binaries suggest that values of
$\alpha \approx 0.1$ may be common in ionized accretion discs (see the review by
\citealt{martin19b}). In this case, disc warps would develop regularly
during X-ray bursts, including many Type I bursts. As X-ray bursts are
frequently detected from over 100 binaries in our Galaxy, detailed
observations with a high-throughput X-ray telescope such as
\textit{STROBE-X} \citep{strobex} would quickly be able to determine
if such values of $\alpha$ occur in NS accretion discs.

The evolution calculations revealed two different types of
warping behaviors that depend on the initial warp given to the
accretion disc. First, we find `pulsations' of warp concentrated in
the inner disc (within $\sim 50$~$r_g$) that grow and dissipate on
timescales of $\sim 1$~s (e.g., Figs.~\ref{fig:type1psi}
and~\ref{fig:imdbpsi}). These pulsations can appear during the plateau
phase of the burst and in the tail when the luminosity is
decreasing. The growth and decay timescale of $\sim 1$~s are
consistent with the radial and vertical diffusion timescales in the
disc (Fig.~\ref{fig:timescales}). The second behavior found in the
calculations is a more slowly growing and larger scale warp that
covers a large fraction of the disc
(Fig.~\ref{fig:halfgammabeta}). This warp steadily grows during the
plateau phase of the burst before fading during the tail.

These two warp behaviors provide possible explanations for the unusual
properties observed in some IMDBs and superbursts. In particular,
consider the achromatic variations observed in the
tail of the IMDBs from 2S~0918-549 \citep{zand05,igb11},
IGR~J17062-6143 \citep{degenaar13} and GRS~1741.9-2853
\citep{barriere15}. In all these systems, the X-ray lightcurve
began showing fluctuations to higher and lower fluxes during the tail
of the bursts. The fluctuations were not periodic, but typically had
timescales of $\sim 1$~s. The scale of the fluctuations was a factor
of a few, with the dips tending to be deeper than the
enhancements. The changes were observed to be achromatic, indicating
that it was not related to Compton-thin
absorption by ionized gas. Crucially, in all cases the fluctuations stopped on
timescales of $\sim 1$~minute and the lightcurves resumed their normal
decline. Many of these properties appear to be consistent with the warp
'pulsations' found in the previous section. In two-dimensions, these
warps will form spiral structures in the disc
\citep[e.g.,][]{pringle97} that could both block
the view of the NS surface (when on the front side) and increase the
reflecting surface (when on the back side). As the surface density
increases in these warps (e.g., Fig.~\ref{fig:type1}), this would
explain the achromatic nature of the fluctuations. The critical
dependences on duration, viscosity and luminosity would also be a natural
explanation for why these fluctuations only appear in luminous IMDBs,
but also their rare and temporary appearance in the burst tails.

It has been noted that the fluctuations only appear in IMDBs that show
evidence for super-expansion, where a shell of material is blown off
the photosphere for hundreds or thousands of km, and the fluctuations
could result from the interaction of the shell with the accretion
disc \citep{igb11}. However, the burst from GRS~1741.9-2853 which demonstrated
fluctuations only had a mild PRE phase with no evidence for
super-expansion \citep{barriere15}. Therefore, the apparent correlation
with super-expansion may be a result that disc warping and a
strong PRE phase are both more likely with high luminosity bursts.

The slower, larger scale warp evolution observed in the calculations
could explain the high reflection fractions and absorbing column
densities found in the spectral analysis of the superbursts from
\foru\ \citep{bs04} and \fouru\ \citep{keek14b}. These two quantities
were found to be enhanced thousands of seconds into the burst, well
into the burst tail. If a warp steadily grew across the disc during
the burst, similar to what is seen in Fig.~\ref{fig:halfgamma}, then
it could have reached a point where sufficient material was brought
into the line of sight to increase the absorbing column density. In
addition, the solid angle subtended by the disc would have grown due
to the warp, strengthening the total reflection signal in the observed
X-ray spectra. \citet{igb11} reported that the \foru\ superburst also
displayed achromatic fluctuations late in the burst (at $t \approx
6300$--$6700$~s), coincident with the increase in column
density found by \citet{bs04}. Given their long duration, it would not
be unexpected for superbursts to have the highest chances of
developing radiation driven disc warps.

\subsection{Caveats and Directions for Improvement}
\label{sub:caveats}
The results presented in Sects.~\ref{sect:criteria}
and~\ref{sect:evolution} show that X-ray bursts
from neutron stars are capable of driving temporary warps in
their surrounding accretion discs. The existence, evolution and
observational consequences of the warps depend on several factors,
including the disc viscosity, any initial warp prior to the burst, the
burst luminosity, and its duration. While compelling, it is important
to recognize the limitations of our proof-of-concept approach, in particular with the
evolution calculations of Sect.~\ref{sect:evolution}, which, due to
the assumption of isotropic viscosity (i.e., $\alpha_2=1/2\alpha$) are
valid only for small warps ($\Psi < 1$) growing in the linear
regime. As noted above, some of our results exceed $\Psi =1$ and
should be treated with caution. Appendix~\ref{app:bndry}
and~\ref{app:resolution} present tests of how sensitive our results are
to the numerical method. Below, we discuss other issues that could be
improved upon in future work.

The development of the equations used in Sects.~\ref{sect:criteria}
and~\ref{sect:evolution} focused on radii far from the inner
boundary of the accretion disc. However, warps often developed or were
transported to small radii, where that approximation would no longer
be valid. Moreover, relativistic effects will impact the gravitational
potential and orbital velocity profile of gas close to the NS. These
effects were also not considered by the calculation. The intrinsic
one-dimensional nature of the model is also a limitation, as the warps
are known to develop a spiral structure \citep{pringle97} due to
differential rotation in the disc. It is also worth noting the
razor-thin nature of the disc assumed in the model. It is possible
that a disc with a realistic thickness, illuminated from both sides,
may have a qualitatively different warping behavior than discussed above.

Clearly, additional numerical simulations of radiated accretion discs
that relax one or more of the above assumptions would be valuable
tests of the results presented here. The long timescale of the disc
warping instability may prove to be prohibitive for detailed global
calculations. Therefore, semi-analytical work, similar to this work,
will remain valuable when considering radiation driven disc warps. The
broad agreement between the analytic estimates of Sect.~\ref{sect:criteria}
and the evolution calculations of Sect.~\ref{sect:evolution} indicate
that the physical mechanisms underlying radiatively driven warps is
likely to be robust and therefore relevant for X-ray bursts and other
explosive phenomena with surrounding accretion discs. 

\section{Conclusions}
\label{sect:concl}
The strong radiation field from X-ray bursts can
significantly alter many properties of the surrounding accretion disc
during the burst \citep[e.g.,][]{degenaar18, fbb19}. In this paper we
demonstrate that bursts can also generate radiatively driven warps in
discs. Disc warps will be easier to produce for luminous, long
duration bursts (Fig.~\ref{fig:criteria}), and therefore we expect signatures of warping to be
more common during IMDBs and superbursts, especially those with PRE
events, which reach the highest luminosities. Additionally, the ability for a warp to grow depends on
the strength of the disc viscosity, with warps being more likely if
$\alpha \ga 0.1$. Given the dependence on viscosity, it is possible that
warps may be a relatively rare or short-lived occurrence during an IMDB
or superburst. However, if evidence for warps is found regularly
during Type I bursts then this would indicate that $\alpha \ga
0.1$. Such an investigation would be well within the capabilities of the
\textit{STROBE-X} mission concept.

We considered the evolution of a burst-driven warp in the linear
regime due to a Type I burst (lasting $25$~s) and IMDB (lasting
$250$~s) and found two different behaviors for the warps, depending on
the initial warp seeding the disc. In some models, rapid ($\sim 1$~s)
pulsations of warp growth and decay were found in the inner regions of
the disc ($r \la 50$~$r_g$) during both the plateau and tail of the
bursts (Figs.~\ref{fig:type1} and~\ref{fig:imdb}). The timescale of
the pulsations were consistent the radial and vertical diffusion times
of the disc. The properties of these stochastic warps appear to be
consistent with the achromatic fluctuations observed in the tails of
several IMDBs \citep[e.g.,][]{igb11,degenaar13,barriere15}, and
provide a natural explanation for their timescales,
appearance and disappearance. In the second scenario, the disc
developed a a more slowly
growing warp covering a large fraction of the disc. The growth of the
warp was proportional to the duration of the burst, implying that
superbursts could develop significant warps over a wide range of
radii. This behavior could explain the large reflection fractions and
absorbing column densities seen late in the superburst spectra from
\foru\ and \fouru\ \citep{bs04,keek14b}.

Although idealized, the calculations presented here show that
radiatively driven accretion disc warps could be a natural consequence
of X-ray bursts. This is yet another illustration of how the
interaction of bursts with accretion discs presents an unique
opportunity to probe accretion processes with a repeatable
experiment. As the growth of the warps depend on the
viscosity acting in the disc, they provide a potentially important
method to observationally constrain the viscous processes within
accretion flows. Future work is needed to extend the proof-of-concept
studies of this paper and predict observationally testable properties
of burst driven disc warps.

\section*{Data Availability}
The data underlying this article will be shared on reasonable request
to the corresponding author.

\section*{Acknowledgements}
The author thanks O.\ Blaes, C.\ Reynolds and J.\ Dexter for helpful
feedback during preparation of this work.




\bibliographystyle{mnras}
\bibliography{refs.bib} 



\appendix
\section{Warp Evolution with Free Inner Boundary Condition}
\label{app:bndry}
\begin{figure}
  \includegraphics[width=0.48\textwidth]{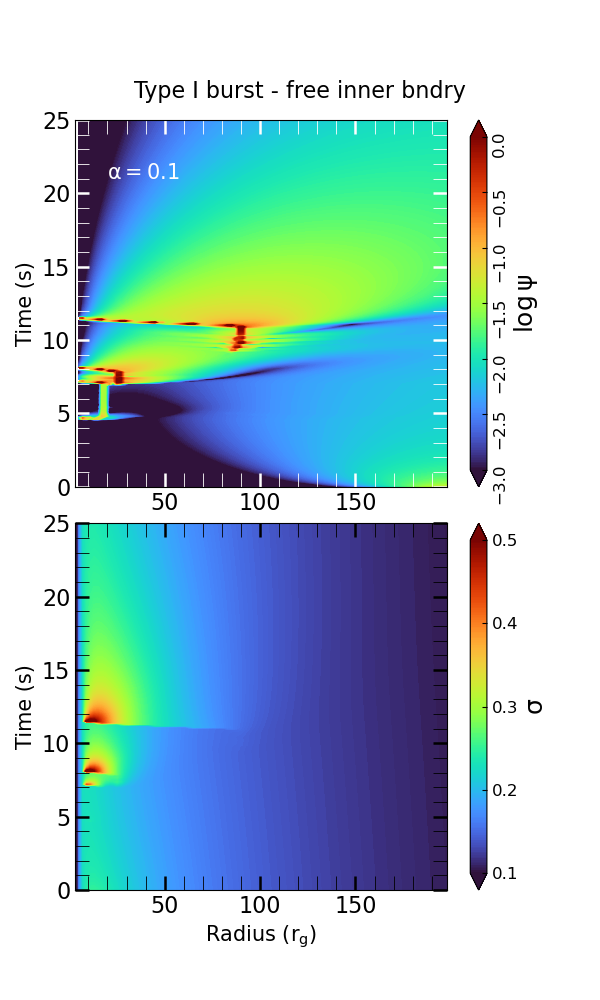}
  \caption{Results of the warp evolution calculation for a Type I
    burst with $\alpha=0.1$ when $\beta$ and $\gamma$ are not set to
    zero at
    the inner boundary (however, the $\partial \bm{\ell}/\partial r=0$
    condition remains at the inner boundary).  This result should be compared with the the
  right-side of Fig.~\ref{fig:type1}.}
  \label{fig:appA}
\end{figure}

One of the assumptions described in Sect.~\ref{sub:numerical} is that
the disc tilt and twist angles, $\beta$ and $\gamma$, are set to zero
at the inner boundary to account for the probable effects of the disc
meeting a boundary layer at the NS surface. There is enough
uncertainty, however, in the details of the transition to
consider the case where $\beta$ and $\gamma$ are not forced to zero at
the boundary. Figure~\ref{fig:appA} shows the results for this
situation for a calculation of a Type I burst impacting a disc with
$\alpha=0.1$. This figure should be compared with the right side of
Fig.~\ref{fig:type1}. While $\beta$ and $\gamma$ are free at the inner
boundary, we still enforce zero torque and $\partial
\bm{\ell}/\partial r=0$ conditions.

The figure shows strong warps developing during the plateau and decay
phase of the burst. Unlike the situation with $\beta=\gamma=0$ at the
inner boundary, warps grow both at $\sim 25$~$r_g$ and also at $\sim
90$~$r_g$. However, pulsations of warp at $\la 1$~s long timescales
still occur at both locations. The disc tilt at the inner edge varies
throughout the burst, reaching a maximum of $11.1$~degrees at
$8.1$~s. The warps become strongly non-linear, with $\Psi=5.2$~at
$11.1$~s and $65$~$r_g$, and $\Psi > 1$ at $t \sim 7$--$8$~s. The
surface densities at $r < 25$~$r_g$ also increase markedly during the
periods of non-linear warps. We conclude that the $\beta=0$ and
$\gamma=0$ boundary condition used in the paper produces a less extreme result that
remains safely in the linear regime. 

\section{Resolution Check}
\label{app:resolution}
\begin{figure}
  \includegraphics[width=0.48\textwidth]{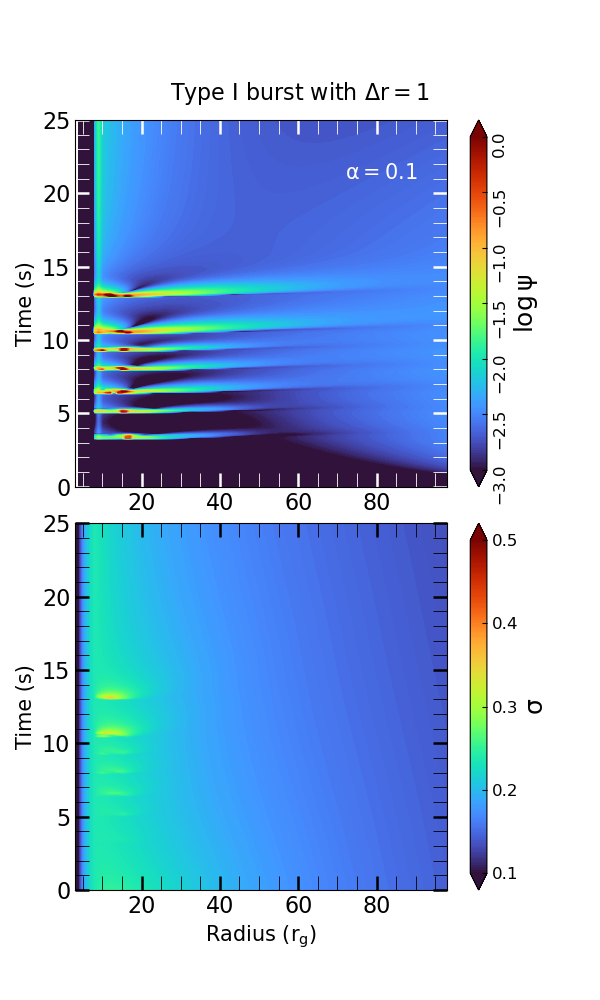}
  \caption{Results of the warp evolution calculation for a Type I
    burst with $\alpha=0.1$ when the resolution of the radial grid is
  doubled to $\Delta r=1$. This result should be compared with the the
  right-side of Fig.~\ref{fig:type1}.}
  \label{fig:dr1}
\end{figure}

The results presented in the paper are calculated on a radial grid
from $3$ to $401$~$r_g$ with $\Delta r=2$~$r_g$. The linear nature of
the grid was chosen because, \emph{a priori}, it was unknown where in
the disc any warp would start to develop. Indeed, the predictions of
Fig.~\ref{fig:criteria}, suggested a warp radius of $\sim 100$~$r_g$,
as was found in Fig.~\ref{fig:halfgamma}. To examine the
effects of the resolution of the grid, in particular on the properties
of the small-scale warp `pulsations' seen in Figs.~\ref{fig:type1}
and~\ref{fig:imdb}, we show in Figure~\ref{fig:dr1} the results from
repeating the $\alpha=0.1$, Type I burst calculation with the
resolution doubled to $\Delta r=1$~$r_g$.
Comparing this result to the one shown in the right-hand side
of Fig.~\ref{fig:type1} shows that warp pulsations are still
predicted during the plateau phase of the burst, but the ones in the
tail do not appear in the $\Delta r=1$~$r_g$ calculation. However, the
pulsations seen in Fig.~\ref{fig:dr1} are more frequent and stronger
during the plateau phase of the burst. The warp peaks are separated by
$\sim 1.5$~s, consistent with the radial diffusion time at the inner
edge of the disc (Fig.~\ref{fig:timescales}). Three of the pulsations
exceed $\Psi = 1$, with the last one reaching $\Psi=2.2$
(corresponding to a $\beta=5.6$~degrees) at
$t=13.1$~s, indicating that the disc warps are more likely to reach
the non-linear regime in higher-resolution simulations. Overall, we
conclude that, while the details may change, rapid pulsations of disc
warp remain a possible outcome of radiatively driven warps by X-ray
bursts. Higher resolution calculations, using, e.g., geometric
spacing, are needed to follow the warps into the non-linear regime and
will be pursued in future work.


\bsp	
\label{lastpage}
\end{document}